\documentclass[twocolumn]{aastex61}
\pdfoutput=1 
\usepackage{amsmath,amstext}
\usepackage[T1]{fontenc}
\usepackage{apjfonts} 
\usepackage[figure,figure*]{hypcap}
\usepackage{graphicx}
\usepackage{nicefrac}
\usepackage{color}
\usepackage{multirow}
\usepackage{tablefootnote}
\usepackage{epsfig}
\usepackage{xcolor}

\shorttitle{Microlensing and Intrinsic Variability of the Broad Emission Lines of Lensed Quasars}
\shortauthors{Fian et al.}

\begin{document}
\title{Microlensing and Intrinsic Variability of the Broad Emission Lines of Lensed Quasars}

\author{C. Fian}
\affiliation{Instituto de Astrof\'{\i}sica de Canarias, V\'{\i}a L\'actea S/N, La Laguna 38200, Tenerife, Spain}
\affiliation{Departamento de Astrof\'{\i}sica, Universidad de la Laguna, La Laguna 38200, Tenerife, Spain}
\author{Eduardo Guerras}
\affiliation{Homer L. Dodge Department of Physics and Astronomy, The University of Oklahoma, Norman, OK, 73019, US}
\author{E. Mediavilla}
\affiliation{Instituto de Astrof\'{\i}sica de Canarias, V\'{\i}a L\'actea S/N, La Laguna 38200, Tenerife, Spain}
\affiliation{Departamento de Astrof\'{\i}sica, Universidad de la Laguna, La Laguna 38200, Tenerife, Spain}
\author{J. Jim{\'e}nez-Vicente}
\affiliation{Departamento de F\'{\i}sica Te\'orica y del Cosmos, Universidad de Granada, Campus de Fuentenueva, 18071 Granada, Spain}
\affiliation{Instituto Carlos I de F\'{\i}sica Te\'orica y Computacional, Universidad de Granada, 18071 Granada, Spain}
\author{J. A. Mu\~{n}oz}
\affiliation{Departamento de Astronom\'{i}a y Astrof\'{i}sica, Universidad de Valencia, E-46100 Burjassot, Valencia, Spain}
\affiliation{Observatorio Astron\'{o}mico, Universidad de Valencia, E-46980 Paterna, Valencia, Spain} 
\author{E. E. Falco}
\affiliation{Harvard-Smithsonian Center for Astrophysics, 60 Garden St, Cambridge, MA 02138, USA}
\author{V. Motta}
\affiliation{Instituto de F\'{\i}sica y Astronom\'{\i}a, Universidad de Valpara\'{\i}so, Avda. Gran Breta\~na 1111, Playa Ancha, Valpara\'{\i}so 2360102, Chile}
\author{A. Hanslmeier}
\affiliation{Institute of Physics (IGAM), University of Graz, Universit{\"a}tsplatz 5, 8010, Graz, Austria}

\begin{abstract}
We study the broad emission lines (BELs) in a sample of 11 gravitationally lensed quasars with at least two epochs of observation to identify intrinsic variability and to disentangle it from microlensing. To improve our statistical significance and emphasize trends we also include 15 lens systems with single-epoch spectra. MgII and CIII] emission lines are only weakly affected by microlensing, but CIV shows strong microlensing in some cases, even for regions of the line core, presumably associated with small projected velocities. However, excluding the strongly microlensed cases, there is a strikingly good match, on average, between the red wings of the CIV and CIII] profiles. Analysis of these results supports the existence of two regions in the broad line region (BLR), one that is insensitive to microlensing (of size $\gtrsim 50$ light-days and kinematics not confined to a plane) and another that shows up only when it is magnified by microlensing (of size of a few light-days, comparable to the accretion disk).  Both regions can contribute in different proportions to the emission lines of different species and, within each line profile, to different velocity bins, all of which complicates detailed studies of the BLR based on microlensing size estimates. The strength of the microlensing indicates that some spectral features that make up the pseudo-continuum, such as the shelf-like feature at $\lambda1610$ or several FeIII blends, may in part arise from an inner region of the accretion disk. In the case of FeII, microlensing is strong in some blends but not in others. This opens up interesting possibilities to study quasar accretion disk kinematics. Intrinsic variability seems to affect the same features prone to microlensing, with similar frequency and amplitude, but does not induce outstanding profile asymmetries. We measure intrinsic variability ($\lesssim$20\%) of the wings with respect to the cores in the CIV, CIII], and MgII lines consistent with reverberation mapping studies.  

\end{abstract}

\section{Introduction \label{intro}}
To date, the primary probe of the geometry and kinematics of the  broad line regions {(BLR) of AGN} has been reverberation mapping (hereafter RM, \citealt{Blandford1982}, see also \citealt{Peterson2006} and references therein) based on the measurement of the lag between the intrinsic variability of the continuum and of the broad emission lines (BEL). This technique has shown that the global structure of the broad line region is consistent with photoionization models \citealt{Bentz2009}. RM results also suggest that there is no common kinematic  structure, with differing sources showing signs of inward, outward, and rotating disk-like velocity structures (e.g., \citealt{Bentz2010}, \citealt{Grier2017a} and references therein). Until now, RM studies have been largely limited to relatively nearby, lower luminosity AGN. Very recently,  new results for individual (\citealt{Shen2016}, \citealt{Grier2017b}) or composite (\citealt{Li2017}) sources ($z\lesssim 1.1$) of intermediate-luminosity  based on the first year of the Sloan Digital Sky Survey RM project, have been presented, mainly for H$\alpha$, H$\beta$, and MgII. However, there are few RM measurements for CIV (\citealt{Kaspi2007}), which is the most easily studied line in quasars with $1.5 \lesssim z \lesssim 2.5$.\\

An alternative means of studying the structure of the BLR is to examine how it is microlensed in gravitationally lensed quasars. In microlensing, the stars in the lens galaxy differentially magnify components of the quasar emission regions, leading to time- and wavelength-dependent changes in the spectra of the images (\citealt{Wambsganss2006}, \citealt{Abajas2002}). The amplitude of the magnification is controlled by the size of the emission region, with smaller source regions showing larger magnifications. Microlensing is now a well-established tool for studying quasar structure, for both individual objects and larger statistical samples (\citealt{Sluse2012}, \citealt{Motta2012}, \citealt{Guerras2013a,Guerras2013b}).\\

In \citealt{Guerras2013a} (see also \citealt{Motta2012}) we compared the emission line profiles of pairs of images from a sample of 13 lensed quasars for which archival single-epoch spectroscopy is available. For single-epoch data some method is needed to separate the effects of microlensing from those of extinction and uncertainties in the magnification produced by the lens galaxy (macro-magnification). In principle, the cores of the emission lines can be used as a reference that is little affected by microlensing. Specifically, the ratio of the line core fluxes of two images to their line wing fluxes gives an estimate of the size of the wing emission regions. \citealt{Guerras2013a} were able to measure wing microlensing magnifications of $\sim$ 0.1 mag, which, under the above hypothesis, led to estimates for the BLR size of  $r_s=24_{-15}^{+22}$ and $r_s=55_{-35}^{+150}$~light-days (90\% confidence)  for the high and low ionization lines, respectively. As expected from ionization stratification, high ionization lines come from a smaller region than low ionization lines, and higher luminosity quasars have larger emission regions. The sizes were somewhat smaller than those found for existing (full line) reverberation studies, as expected from measuring the size of the line wings. Using similar techniques, \citealt{Guerras2013b} have found a strong microlensing signal (up to $\sim$ 0.6 mags) in the UV pseudo-continuum emission from  FeII and FeIII, indicating that the region emitting this pseudo-continuum is quite small, with a size comparable to that of the accretion disk determined from microlensing of the continuum.\\

  As a qualitative leap forward from the static single-epoch analysis, here we study the variability of the emission lines by comparing spectroscopy from at least two different epochs. The response to microlensing over time of different kinematic regions (corresponding to emission lines of different origin or to different wavelength regions in a given emission line)  contains a large amount of information that can improve our understanding of the BLR structure and kinematics. The analysis may be complex, for in a microlensing event the source structure and the pattern of microlensing magnification are convolved, but the expectations are promising.\\

On the other hand, the comparison of two epochs of spectroscopic observations is necessary to study intrinsic quasar variability that may be of great interest by itself and in comparison with microlensing. In particular we would like to compare the frequency and amplitude of both sources of variability and determine whether they affect the same kinematic regions or not. This last question may be very interesting when cross-checking the results obtained from reverberation mapping and microlensing.\\

Spectroscopic monitoring of at least two epochs is also useful to separate intrinsic variability and microlensing. In principle, at a given epoch, intrinsic variability should affect all the images of a lensed quasar in the same way, whereas microlensing may induce differences between the spectra of different images. However, the time delay in the arrival of the light from two lensed images can complicate this simple scheme by inducing differences between images  which may arise from intrinsic variability. Although there are reasons to suppose that the contamination by intrinsic variability modulated by the time delay is weak \citep{Guerras2013b}, a second epoch of spectroscopic observations can allow us to untangle the causes of variability. If the different images change in the same way, the origin of the variability must be intrinsic but if one of the images exhibits differences with respect to the others we may expect that it is affected by microlensing.\\

Finally, the changes in the spectra of an image between two epochs are completely unaffected by extinction and uncertainties in the lens model and hence potentially provide independent estimates of the microlensing of the core and wings. This enables us to make independent estimates of the sizes of the two regions rather than (to be precise) of the differences in the sizes of the two regions measured when only a single epoch is available.\\

Obviously, all these positive prospects for the study of BEL variability are affected by the quality of the data, which, in our case include several old spectra available in the literature, as described in \S 2. \S 2 also includes the data analysis. \S 3 is devoted to the discussion of the results in the context of BLR structure and kinematics. Finally, the main conclusions are summarized in \S 4.\\

\section{Data Analysis and Results}

\subsection{Data}

{We collected from the literature rest frame UV spectra of lensed quasars obtaining a sample of 26 lensed quasars of which 11 are observed in (at least) two different epochs. Information about the observations and references are summarized in Table \ref{1}.}\\

As commented in \S1, one of the motivations to consider a second epoch spectroscopy  is the possibility of comparing the intensity of a given spectral feature (the wing of the CIII] line, for instance) of the same image ($A$) in two epochs ($I$ and $II$), to study differences of the type $(A^I-A^{II})_{wing(CIII])}$. This difference is free from extinction or inaccurate macro-modeling effects.  However, it is not free from instrumental problems such as slit misalignment or errors in the  calibration of the instrumental response with wavelength, which, unfortunately, are likely present in the available data. Thus, we need a non-changing reference spectral feature to scale the spectra. As a first step (\S \ref{matching}) we consider the usual assumption of single epoch based studies that the cores of CIV, CIII] and MgII (defined by the flux within a narrow interval of $\pm$6\AA\ centered on the peak of the line) arise from a substantially larger area than the wings ($\sim$ 30\AA\ wavelength intervals on either side of the emission line peak) and are consequently little affected by microlensing (see  \citealt{Guerras2013a} and references therein). Specifically, we start using the cores of the CIV, CIII], and MgII lines as references to study the wings of these lines and other weaker emission lines and blends in the CIV to MgII wavelength range. Next (\S \ref{cores}), we discuss the hypothesis of unchanging cores by studying the variability of the cores of CIV and CIII], taking MgII as reference.

\subsection{Variability Taking as Reference the Cores of the CIV, CIII], and MgII Emission Lines\label{matching}}

{Figures \ref{spectra1} to \ref{spectra4} show, for 11 lens systems with at least two epochs of observation (Table \ref{1}), the core-matched spectra in the wavelength regions around the CIV, CIII], and MgII lines (see Appendix \ref{AA} for details about the core-matching analysis). The spectra of 26 lens systems (11 with at least two epochs of observation and 15 with single-epoch spectra), split into two wavelength regions from CIV to CIII] and from CIII] to MgII, are shown in Figures \ref{iron1} to \ref{iron5}}. 
Under the assumption that the line core arises from a significantly larger region than the wings, the $ (F_{1core}/F_{2core})/(F_{1wings}/F_{2wings})$ ratio of the line core fluxes of two images, $F_{1core}/F_{2core}$, to the line wing fluxes of these two images,  $F_{1wings}/F_{2wings}$, yields a measurement of the size of the high velocity wing emission regions. The computed magnitude differences between images with S/N ratio greater than 1.5 (see Appendix \ref{AA} for the details of the computation) are given for each epoch in Tables \ref{1.5em_images} (CIV, CIII], and MgII) and \ref{1.5o_images} (other considered emission line features). Similarly, the magnitude differences between epochs for each image are presented in Tables \ref{1.5em_epochs} (CIV, CIII], and MgII) and \ref{1.5o_epochs} (other emission line features). To qualify the magnitude differences between epochs or/and images as candidates for intrinsic variability or microlensing we have used the following criteria: (i) the S/N ratio should be greater than 2, (ii) any difference between images is considered a candidate for microlensing, (iii) we consider as a candidate for intrinsic variability a difference between two epochs when it is present in at least two images, (iv) when neither (ii) nor (iii) apply we consider that we have insufficient information to qualify the difference, although intrinsic variability may be more likely (partial evidence of intrinsic variability). The resulting classification of the differences between spectra corresponding to several emission line features in the CIV to MgII wavelength range is included in Tables \ref{1.5em_images}, \ref{1.5em_epochs}, \ref{1.5o_images}, and \ref{1.5o_epochs}.\\ 

The first result that can be inferred from Figures \ref{spectra1} to \ref{iron5} as shown in Tables \ref{1.5em_images} to \ref{1.5o_epochs} is that the MgII line is weakly affected by either microlensing or intrinsic variability with no significant changes (S/N $\gtrsim 2$) in the selected windows.\footnote{Except for a difference between images D and B in the blue wing of HE 0435-1223 at epoch II.} A similar result can be derived for the  CIII] line.\footnote{With only two exceptions;  B-A at epoch I in HE 0435-1223 and C-B at epoch I in WFI J2033-4723.} \\

In the case of CIV the presence of microlensing or intrinsic variability is more the rule than the exception, particularly in the extreme red wing. In Figure \ref{mean+sigma} we show the mean+$\sigma$ spectra of the CIV emission line for each system to emphasize the variability in the wings. There is evidence, at the 2$\sigma$ level, of intrinsic variability within the defined integration windows in one image of Q 0142-100 and in HE 1104-1805.  There is also evidence of intrinsic variability in the extreme red wing of CIV, specifically in the shelf-like feature blueward of He II, around $\lambda1610$.\footnote{In Q 0142-100, QSO 0957+561, HE 1104-1805 and HE 2149-2745} Microlensing is detected in CIV, at the 2$\sigma$ level, in the integration windows\footnote{In  SDSS J1004+4112, HE 0435-1223, and SDSS J1339+1310.} as defined in Figure \ref{spectra1} and in the shelf-like feature at $\sim \lambda1610$.\footnote{It may be strongly microlensed ($\gtrsim 2\sigma$) in  HE 0435-1223, SDSS J1004+4112 and SDSS J1339+1310.}  The origin of this feature is uncertain (e.g., \citealt{Fine2010} and references therein). It has been interpreted as an (extreme) CIV red wing or another species. In most of our objects\footnote{Q 0142-100, HE 0435-1223, QSO 0957+561, HE 1104-1805, and SDSS J1339+1310} microlensing or intrinsic variability seems to affect, in a more or less smooth way, the whole red wing. {However, the absence of a blue counterpart of this feature would need further explanation}. On the other hand, in SDSS J1004+4112 (and perhaps HE 2149-2745),  a feature at $\sim \lambda$1610 shows up very distinctly{, which supports the hypothesis of an unidentified species.} Apart from this feature, in the pseudo-continua between CIV and CIII], and between CIII] and MgII, there are several lines and blends showing evidence of intrinsic variability and microlensing at the 2$\sigma$ level (see Appendix \ref{AB}), that very noticeably affect the complex formed by the HeII emission line, the OIII]/AlII blend, and the underlying  pseudo-continuum in the CIV-CIII] wavelength range\footnote{Other lines in this wavelength range, such as AlIII and SiIII], are also affected by microlensing but their study is hampered because they are strongly blended with CIII].}, and the FeII and FeIII blends in the CIII]-MgII range. \\

According to the criteria explained in Appendix \ref{AA} (see Tables \ref{1.5em_images} to \ref{1.5o_epochs}), we can consistently separate most of the observed systems in terms of intrinsic variability or microlensing. There are four objects clearly dominated by microlensing (SDSS J0806+2006, FBQS J0951+2635, SDSS J1004+4112, and SDSS J1339+1310) and five objects in which intrinsic variability prevails (Q 0142-100, HE 1104-1805, WFI J2033-4723, HE 2149-2745, and HE 0047-1756; in the last object there is evidence of microlensing in epoch I). The differences observed in QSO 0957+561 may be explained by intrinsic variability combined with the large time-delay between the images of this double plus a possible contribution from microlensing. Finally,  HE 0435-1223, shows both intrinsic variability\footnote{Strong in the shelf-like feature at  $\sim \lambda$1610, He II, Fe II, and Fe III, and weaker in C IV.} and microlensing of relatively lower  amplitude.\footnote{Mainly in B-A (epoch IV in C IV, shelf-like feature at $\lambda 1610$ and FeIII).} \\

In Figures \ref{histo_images} and \ref{histo_epochs} we present, for the different species, histograms of microlensing magnification and intrinsic variability. We have overlaid the corresponding Gaussian kernel density estimates of the probability density functions (PDFs)  to show the impact of errors in the individual measurements. According to these figures, within the limitations of the size of the sample of lens systems and of the data quality, intrinsic variability and microlensing seem to affect the same spectral features with similar strength. The impact of microlensing (and of intrinsic variability) looks similar for the shelf-like feature at $\sim \lambda 1610$, Fe II and III and significantly smaller for CIV. Comparing with microlensing data of the continuum we can say that the former features exhibit microlensing values similar to those typical of the optical continuum ($|\Delta m_{opt}| \sim 0.3\,\rm mag$, see \citealt{Jim2015}), and that, in extreme cases, some spectral features (FeIII$\lambda\lambda$2039-2113, for instance) can undergo microlensing magnifications close to those typical of the X-ray continuum  ($|\Delta m_{Xray}| \sim 1\,\rm mag$, see \citealt{Jim2015}).\\

Moreover, we obtain rms intrinsic variabilities of the wings with respect to the cores of 18\%, 12\%, and 15\% for the CIV, CIII], and MgII lines, respectively. This implies that the intrinsic variability affects the wings and core with different intensity, and/or that there is a delay between both. This difference is observed in RM studies with resolution in velocity and is the basis for studying the kinematic structure of the BLR (e.g., \citealt{Bentz2010,Grier2017a} and references therein). Typical rms variabilities for CIV and CIII]  of about ($13\pm9$)\% have been found by \citealt{Kaspi2007}. For the MgII line, an rms variability of about ($19\pm 3$)\% can be inferred from the reduced sample of sources of \citealt{Shen2016}. In spite of the reasonable agreement between these quantities and our results, the extent of the comparison is limited by the heterogeneity  of the samples (in luminosity, redshift, and timescales) and by the fact that \citealt{Kaspi2007} and \citealt{Shen2016} are measuring the variability of the whole line.

\subsection{Variability of the Cores of the CIV and CIII] Emission Lines taking MgII as Reference\label{cores}}

One important consequence of the previous section is that there is little differential microlensing between the core and wings of MgII (the same result also applies to the red wing of CIII]). This result looks, in principle, consistent with the relatively small range of velocities corresponding to the wings of MgII (as compared with CIV). However, this conclusion should be regarded with caution for, as we shall see later, the correspondence between velocity channel and microlensing impact is not simple. On the other hand,  there are several systems (SDSS J1004+4112 and SDSS J1339+1310, in particular) in which microlensing is strongly affecting the wings of CIV (with respect to the core) without traces of this microlensing in the wings of MgII. Consequently,  it seems very unlikely that microlensing is present in MgII but does affect the core and the wings with the same strength. Thus, it is reasonable to assume that the entire MgII line is weakly sensitive to microlensing and to use this line as a reference to match the spectra of an image at different epochs. Therefore we circumvent the problems originating from inaccuracies in the alignment of the slit and in the calibration of the spectra, and thus obtain a measurement of intrinsic variability unaffected by extinction or macro-magnification modeling.\\

\subsubsection{Core Microlensing}

Let us start the core microlensing study (taking MgII as reference) by computing, for each available epoch, differences of the type $(B-A)_{core(CIII])}-(B-A)_{core(MgII)}$, which we expect to be less affected by instrumental and calibration problems. In principle this difference is sensitive to the differential microlensing between the cores of both emission lines, to differential extinction, and to differences in intrinsic variability between both lines during the time delay between A and B. If we assume that the MgII line is fairly insensitive to microlensing, this difference basically measures extinction and the impact of microlensing in the CIII] core. The average of the absolute value of these differences is $0.09\pm0.08$ mag (68\% confidence interval) indicating that there is little differential microlensing between CIII] and MgII (as shown in Table \ref{extinction}). \\

We repeat the same calculation for the CIV core but now using the core of CIII] as reference, $(B-A)_{core(CIV)}-(B-A)_{core(CIII])}$, in order to prevent the larger extinction impact of a direct comparison with MgII, obtaining a mean value for the absolute differences of $0.12\pm0.11$ mag (68\% confidence interval, see Table \ref{extinction}). Taking into account that the typical rms uncertainty in the determination of a single B-A difference is around 0.05 mag and other possible sources of uncertainty of the matching process, we can conclude that microlensing should have, on average, little impact in the cores of MgII, CIII], and CIV lines.\footnote{For the same reason, differential extinction may have on average only a marginal impact between MgII and CIII], and between CIII] and CIV although it is clear that extinction can play an important role in some of the outliers in the histograms of core microlensing (not shown here).} However, it is important to notice that in the two strongest cases of microlensing in the CIV line, if we match the CIII] lines, the core of CIV is affected by microlensing with maximum amplitudes of $0.23\pm0.07$ mag (SDSS J1004+4112) and $0.16\pm0.03$ mag (SDSS J1339+1310). \\

\subsubsection{Core Intrinsic Variability \label{core}}

We can now try to compare a given image in two different epochs by computing the differences $(A^I-A^{II})_{core(CIII])}-(A^I-A^{II})_{core(MgII)}$ or $(A^I-A^{II})_{core(CIV)}-(A^I-A^{II})_{core(CIII])}$. These types of quantities are supposed to be free from extinction, and are affected by intrinsic variability and microlensing. After our previous conclusion about the low impact of microlensing in the cores of CIV, CIII], and MgII, we can reasonably think that they will mainly measure intrinsic variability. These quantities can nevertheless be strongly affected by uncertainties in the calibration of the spectral response at two different epochs. We have computed the absolute values of the differences between the cores of CIII] (with respect to MgII) and CIV (with respect to CIII]) to obtain averages of $0.24\pm0.21$ mag and $0.29\pm0.25$ mag for the CIII] and CIV cores, respectively (see Table \ref{corevar}). The scatter is too high to establish any reasonable comparison with other measurements. Nevertheless, we should take into account that, owing to the heterogeneity of the instrumentation used to obtain the data, errors of 20\% in the spectral response in the observed wavelength range cannot be discarded. Therefore, we cannot be sure about the real impact of intrinsic variability on the cores, and the above values should be interpreted as upper bounds. \\

As a consequence of the analysis of this section, the microlensing results of \S \ref{matching} based on the hypothesis of core matching in a given epoch are reliable, whereas it cannot be excluded that the results concerning intrinsic variability of the wings might be somewhat affected by core intrinsic variability. 

\subsection{Microlensing Variability}

We have identified differences in a given image between two epochs that can be consistently attributed to microlensing variability in the four systems dominated by microlensing (SDSS J0806+2006, FBQS J0951+2635, SDSS J1004+4112, SDSS J1339+1310) and, at a lower amplitude, in HE 0435-1223. In all these systems, microlensing variability is clearly noticeable in several Fe III and Fe II blends. The well studied case of SDSS J1004+4112 (\citealt{Richards2004,Gomez2006,Motta2012,Fian2016} and references therein) presents outstanding examples of variability induced by microlensing in all the high ionization lines. Note the high asymmetry of the effect, dominant in the blue part of the lines (see Fig. 7). In contrast, in SDSS J1339+1310 the enhancement of CIV is clearly asymmetrical toward the red. From \citealt{Mosquera2011}, we derived the effective transverse velocity for SDSS J1004+4112 and SDSS J1339+1310, and made two rough estimates of the distance moved by the accretion disk relative to the magnification pattern during the time elapsed between epochs of observation ($\sim$ one year in SDSS J1339+1310 and $\sim$5 years in SDSS J1004+4112). For SDSS J1339+1310 the distance traveled in the source plane is too small (less than 0.6 light-days) to see variability in both wings. In SDSS J1004+4112 we have obtained more interesting results, owing to a larger displacement of the source ($\sim$4 light-days), leading to variations in both wings. If we compare epochs I and II we can see that the magnified blue wing fades while the red wing enhances. This supports that the separation between the approaching and receding parts of the microlensed region of the BLR is of about a few light-days.\\

In several cases, mainly affecting the shelf-like feature at $\sim\lambda 1610$ and several blends of Fe II and Fe III,  the impact of microlensing and microlensing variability (see Tables \ref{1.5em_images} to \ref{1.5o_epochs}) is $|\Delta m|\gtrsim 1$ mag, comparable to  the typical microlensing magnification amplitudes observed in the X-ray continuum.\\

It is important to notice that the spectral features mainly affected by intrinsic variability are also the CIV wings, some blends of Fe II and Fe III, and the shelf-like feature at $\sim\lambda 1610$. In contrast with microlensing, we have not observed marked asymmetries induced by intrinsic variability in the line profiles. This result supports (within the statistical significance of our relatively limited sample) the hypothesis that the small region sensitive to microlensing is not systematically affected by extinction, beaming, or any other mechanism that may selectively enhance one part of it. Consequently, the asymmetric enhancements observed, either in the blue or red parts of microlensed line profiles, likely originate from an anisotropic distribution of microlensing magnification in the source plane. \\

\section{Discussion}
\subsection{Microlensing and Kinematics}

The main result of the previous sections is that the cores of CIV, CIII], and MgII, and the wings of the last two lines are only weakly affected by microlensing. However, the wings of CIV can be strongly affected. {In other high ionization emission line features, such as the FeIII$\lambda\lambda$2039-2113 blend and the shelf-like feature at $\sim \lambda1610$, the whole feature may be globally affected by high magnification microlensing.}\\

To discuss these results with more detailed kinematic information than the separation between core and wings, in Figure \ref{line_velocity_matched} we have overlapped the averaged line profiles corresponding to CIV, CIII], MgII, and FeIII$\lambda\lambda$2039-2113. The wings of MgII correspond to relatively low velocities as compared with other lines. This is consistent with the weak impact of microlensing on this line. Regarding CIII] and CIV, the red part of these lines match very well, except at the lowest intensity level, where the shelf-like feature at $\sim \lambda1610$ is present in CIV (the blue side of CIII] is blended with AlIII, SiIII, and FeII, and no reasonable comparison can be made). In spite of this remarkable kinematic coincidence, the CIV line can be strongly affected by microlensing, whereas CIII] seems to be rather insensitive to this effect (see, for instance, the cases of SDSS J1004+4112 and SDSS J1339+1310). In addition, they have a different degree of ionization (CIV is of high ionization and CIII] of low ionization). These results suggest that both lines are mostly generated in the same region, but that there is also a contribution, exclusive to the CIV line, from emitters located in a region small enough to be strongly affected by microlensing. As far as a zero-velocity contribution from the emitters in this region can be expected, the core of the CIV line may also undergo some microlensing at a lower level of amplitude, compatible with our results in \S \ref{core} (about 20\% of microlensing magnification in the cores of SDSS J1004+4112 and SDSS J1339+1310).\\

Even if FeIII$\lambda\lambda$2039-2113 is the iron spectral feature least contaminated by other species, it is still a blend of many FeIII single emission lines \citep{Vestergaard2001}, and its kinematic interpretation is not straightforward. A simple fit to the blend of the average spectrum based on the sum of Gaussians of the same FWHM (considering the nominal wavelengths and strengths of the single FeIII lines of this blend, \citealt{Vestergaard2001}) results in a kinematic FWHM of about 9400 km/s, significantly greater than that corresponding to CIV. In addition, the strength of microlensing in the FeIII$\lambda\lambda$2039-2113 blend can, in some cases, be greater than in the optical continuum, comparable even to that of the X-ray regions.\\

In Figure \ref{m_vs_fwhm} we represent the average amplitude of microlensing with respect of the line broadening for FeIII, CIV, CIII], and MgII. There is a global trend relating high microlensing with line broadening. Notice the high differential microlensing of CIV with respect to CIII], even when both lines have close FWHMs, revealing the existence of the small region prone to microlensing not contributing to CIII]. \\

\subsection{Microlensing and Size\label{size}}

Single-epoch microlensing measurements can be used to estimate (or constrain) the size of the emitting regions (e.g., \citealt{Guerras2013a}). For the size calculations we will treat each event as a single epoch event. From the microlensing magnification corresponding to all the lens image pairs with more than one epoch of observation we compute the joint microlensing probability, $P(r_s)$, of obtaining an average estimate of the size, following the steps described in  \citealt{Guerras2013a},
\begin{eqnarray}
P(r_s)=\prod\limits_{i} P_i(r_s), \\
P_i(r_s)\propto e^{-\frac{{\chi_i^2(r_s)}}{2}},\\
\chi_i^2(r_s)=\sum\limits_{\alpha_i}{\sum\limits_{\beta_i < \alpha_i}\left(\frac{{\Delta m^{obs}_{\beta_i\alpha_i}-\Delta m_{\beta_i\alpha_i}(r_s)}}{\sigma_{\beta_i\alpha_i}}\right)^2},
\end{eqnarray}
where $\Delta m^{obs}_{\beta_i\alpha_i}$ is the observed differential microlensing magnification between images $\alpha$ and $\beta$ of system $i$ and $\Delta m_{\beta_i\alpha_i}(r_s)$ is the differential microlensing magnification predicted by the simulations for a given value of $r_s$. The simulations are based on $3000\times 3000$ pixel
magnification maps, spanning $600\times600$ light-days$^2$ on the source plane, obtained using the Inverse Polygon Mapping method \citep{Mediavilla2006,Mediavilla2011}. The general characteristics of the magnification maps are determined (for each quasar image) by the local convergence $\kappa$ and the local shear $\gamma$, which were obtained by fitting a singular isothermal sphere with an external shear (SIS+$\gamma_e$) that reproduce the coordinates of the images (\citealt{Mediavilla2006}). We have assumed a mean stellar mass of $0.3M_\odot$. To simulate the effect of the finite source we have convolved the magnification maps with 2-D Gaussian profiles of sigma $r_s$, logarithmically spanning an interval between 0.2 to 120 light-days. Sizes are converted to half-light radius multiplying by 1.18, $R_{1/2}=1.18r_s$. \\

The resulting joint likelihood function can be seen in Figure \ref{PDF_lines}. From Figure \ref{PDF_lines} we can estimate a size for the region emitting the low ionization emission of $50.3_{-14.0}^{+30.4}\sqrt{M/0.3 M_\odot}$ light-days. This result is in good agreement with the estimates by \citealt{Guerras2013a}. 
It is also in agreement, within the uncertainties, with the RM size estimates for MgII of the two SDSS-RM sources of highest luminosity in the comparatively low luminosity sample of \citealt{Shen2016} (object 101, $\tau=36.7_{-4.8}^{+10.4}$ days; object 589, $\tau=34.0_{-12.7}^{+6.0}$ days). Finally, our measurement matches very well the \citealt{Bentz2013} size--luminosity relation based on H$\beta$ RM.\\

In principle, we could now repeat the same procedure with the high ionization lines, starting with CIV. However, from our previous kinematic analysis, we know that the CIV microlensed line profile is likely a combination of emission coming from the large region weakly sensitive to microlensing, where CIII] and MgII originate, and from another small region that shows up in the line profile only when microlensing is present. Moreover, we can suspect that the proportion of the contributions from the two regions to the line profile changes with wavelength. Thus, we should refine the previous approach taken in \citealt{Guerras2013a,Guerras2013b} in which the core and wings were supposed to come exclusively from one of the regions. Instead of this approach, which is complex and needs some modeling (we defer it to future work), we are going to suppose that the spectral features with the highest microlensing magnification arise exclusively from the small region susceptible of microlensing. From the likelihood function corresponding to the observed microlensing (see Figure \ref{PDF_other}) we infer (using a logarithmic prior for the size) a size of $4.1_{-0.8}^{+0.8} \sqrt{M/0.3 M_\odot}$ light-days for the $\sim \lambda$1610 feature at the red shelf of CIV, 
comparable to the optical continuum. In Figure \ref{PDF_fe3} we show the PDFs corresponding to three different blends of Fe III ($\lambda\lambda$1978-2018, $\lambda\lambda$2039-2113, and $\lambda\lambda$2386-2449). The estimated average size, $11.3_{-4.0}^{+5.0} \sqrt{M/0.3 M_\odot}$ light-days, is consistent within the errors with the sizes measured for the red shelf feature at $\sim \lambda$1610.
  Actually, the highest microlensing values are measured in Fe III but the presence of several systems with low microlensing amplitudes in the subsample of systems with more than one epoch of observation (used to compute the PDFs), has made slightly larger the size of Fe III. We obtain slightly larger sizes for the emission regions if we take into account the rescaling of the radii by the luminosity of the systems ($r_i = r_0 \sqrt{L_i/L_0}$).\\
  
In Figure \ref{PDF_fe2} we finally present the PDFs corresponding to several blends of FeII ($\lambda\lambda$1705-1730, $\lambda\lambda$1760-1800, $\lambda\lambda$2158-2197, $\lambda\lambda$2209-2239, $\lambda\lambda$2261-2364, $\lambda\lambda$2460-2564, and $\lambda\lambda$2596-2645). The main result that can be inferred from this figure is the different sensitivity to microlensing, which implies different sizes. The microlensing based size of the $\lambda\lambda$2158-2197 blend is $5.2_{-2.2}^{+1.8} \sqrt{M/0.3 M_\odot}$ light-days, comparable to the sizes of the Fe III emitting regions. In contrast, the other blends seem to arise from  regions of significantly larger size.

\subsection{Structure and Kinematics of the BLR}

From the impact of microlensing,  which we have separately studied in the core and wings of the CIV, CIII], and MgII emission lines, we can attempt to broadly outline a basic relationship between kinematics and structure in the BLR. In the first place, both kinematic regions, core and wings, seem to be little affected by microlensing in CIII] and MgII, indicating that these low ionization lines (as has been usually assumed) arise from a large region, with a lower limit in size of about 50 light-days according to  microlensing estimates. The absence of a central dip in any of the cores of the line profiles (as in most quasars and AGN; see \citealt{popovic2004}) very likely indicates that the motion of the emitters contributing to the core is not confined to a plane \citep{Mathews1982}. The average line profiles of CIII] and CIV match very well (at least in the unblended red part), thereby indicating that both arise mainly from the same region. However, the resemblance between the line profiles is broken by the changes induced by microlensing that reveal the existence of a second region that only shows up in the presence of microlensing. This region contributes to the CIV line (with high strength to the wings and a lower amplitude to the core) but not to CIII] or MgII. According to the high impact of microlensing in other high ionization lines and blends studied, these features also arise from this region whose size, according to microlensing estimates, would be a few light-days. This and the high velocities involved make it natural to identify the relatively small region with (part of) the accretion disk. In some cases (FeIII $\lambda\lambda$2039-2113 for instance) the large microlensing magnifications and the high velocities involved support the hypothesis that the emitters may arise from an inner region of the accretion disk. \\

Microlensing provides estimates of the emitting region sizes, which, combined with the Doppler broadening of the emission lines, should help us to study the kinematics of the BLR and the mass of the central black hole, $M_{BH}$, in a similar way as with RM. It is common in RM studies to suppose that the broadening of the lines (FWHM or $\sigma$) is related to the mass of the central BH through the virial theorem according to 
\begin{equation}
M_{BH}\simeq 9.8 \times 10^7 M_\odot\ f \left(\frac{R_{BLR}}{5{\rm\, light\, days}}\right) \left( {\frac{\Delta v_{FWHM}}{10000 {\rm \, km\, s^{-1}}}}\right)^2.
\label{virial}\end{equation}
When we apply this relationship to CIV emission lines with $\Delta v_{FWHM} \sim4700\rm\, km\, s^{-1}$ (corresponding to the average CIV line obtained from our sample), and $R_{BLR}\sim 50.3_{-14.0}^{+30.4}$ light-days (corresponding to the large BLR region little affected by microlensing), we obtain  $M_{BH}\sim 4_{-1.2}^{+2.4} \times 10^8 M_\odot$, for $f=2$, which is a reasonable result for the bright quasars of our sample (\citealt{Mosquera2013}). A consistent result, $M_{BH}\sim 3.9_{-1.4}^{+1.8} \times 10^8 M_\odot$ ($f=2$), is obtained by considering the FeIII $\lambda\lambda$2039-2113 blend of size  $R_{BLR}\sim 11.3_{-4}^{+5}$ light-days and velocity $\Delta v_{FWHM} \sim9400\rm\, km\, s^{-1}$. The coincidence between both estimates, indicates that microlensing-based sizes are in agreement with the hypothesis of virialized kinematics.\\

Microlensing may, in addition, give more precise information relating velocity and size by considering not the mean properties of the line as a whole but discrete velocity bins in the emission line. However, Eq. \ref{virial} cannot be directly applied to a velocity bin. Its direct application, considering Gaussian sources, to velocity bins in the range 1000 to 10000 km/s (in the cases of SDSS J1339+1310 and SDSS J1004+4112), leads to microlensing sizes so small as compared with those inferred from RM, that the obtained central black hole masses would be unexpectedly low.  Consequently, a kinematic model that describes the geometry of the region contributing to each velocity interval is needed, to simulate the impact of microlensing in this region. On the other hand, as discussed above, we should also take into account that the CIV line profile has contributions from two different regions, but that one of them is much more sensible to microlensing than the other.\\

There are other high ionization lines which seem to arise exclusively from the small region sensitive to microlensing that, owing to the high microlensing magnifications observed, we have identified with the accretion disk, even with its inner regions. A study based on these lines is not straightforward, as they usually form blends and the S/N ratio of the available observations is not sufficient. In any case, an interesting conclusion is that the broad emission lines of these species may be used to study the kinematics of quasar accretion disks.\\

\section{Conclusions}

{We have analyzed the BEL of a sample of 11 gravitationally lensed quasars with at least two epochs of observation. We have studied, in most cases, up to 11 different spectral features (emission lines or blends)  between the CIV and MgII lines. Although it is limited, the temporal sampling available has allowed us to identify intrinsic variability and to classify the differences between pairs of spectra as candidates for intrinsic variability or microlensing. The main conclusions are the following:\\

1 - We can consistently separate a group of four systems dominated by microlensing (SDSS J0806+2006, FBQS J0951+2635, SDSS J1004+4112, and SDSS J1339+1310) and another group of five objects in which intrinsic variability prevails (Q 0142-100, HE 1104-1805, WFI J2033-4723, HE 2149-2745, and HE 0047-1756). The case of QSO 0957+561 may be explained by intrinsic variability combined with the time-delay between the images plus a possible contribution from microlensing. Finally,  HE 0435-1223, seems to be a hybrid case with both microlensing and intrinsic variability present. \\

2 - We study the effects of microlensing and intrinsic variability in the core of the lines (which have been considered unchanging in single epoch based studies). On average, we measure a weak microlensing effect in the cores of CIII] ($\lesssim0.09\pm0.08$ mag with respect to MgII) and CIV ($\lesssim0.12\pm0.11$ mag with respect to CIII]), although in the two strongest cases of microlensing, the core of CIV is significantly affected ($0.23\pm0.07$ mag in SDSS J1004+4112 and $0.16\pm0.03$ mag  in SDSS J1339+1310). Taking the cores as reference,  we find that the wings of MgII and CIII] are not significantly affected (at the $2\sigma$ level) by either intrinsic variability or microlensing. On the other hand, the wings of CIV and the other spectral features analyzed ($\lambda$1610 shelf-like feature, HeII, the OIII]/AlII blend, AlIII, SiIII,  FeII, and FeIII) show strong changes. These results basically confirm the existence of two distinct regions suggested in single epoch based studies,\footnote{Which could neither remove intrinsic variability nor be used to study the impact of microlensing on the cores.} one large and insensitive to microlensing and other small and prone to microlensing,  but with a significant nuance: the small region also contributes to the core of the high ionization lines, although it shows up only in presence of strong microlensing. We have also analyzed core intrinsic variability and obtained estimates of $\lesssim0.24\pm0.21$ mag  for CIII] (with respect to MgII) and $\lesssim0.29\pm0.25$ mag for CIV (with respect to CIII]). Owing to the presence of systematic instrumental effects, these values should be regarded as upper limits on the intrinsic variability of the cores.\\

3 - There is evidence of microlensing variability in the four systems dominated by microlensing and in the hybrid case, HE 0435-1223. Owing to the changes in some spectral features (CIV wings mainly), strong microlensing variability can induce very noticeable asymmetries in the line profile shape. Intrinsic variability affects the same spectral features with similar strength, although no outstanding evidence of asymmetry associated with intrinsic variability has been detected. These results support the hypothesis that the small region susceptible to both intrinsic variability and microlensing is intrinsically symmetrical (i.e., not differentially obscured by dust or magnified by relativistic beaming), and that the asymmetry induced by microlensing in the line profile is related to the anisotropic spatial distribution of microlensing magnification at the source plane.\\

4 -  The relative impact of microlensing indicates that the MgII and CIII] emission lines arise from a region $\sim$ 50 light-days in size, in good agreement with RM studies.  The kinematic coincidence, in the absence of microlensing, between CIII] and CIV supports the hypothesis that a large part of the CIV line also arises from this large region. As the cores of the lines show no central dip, the hypothesis of motion not confined to a plane is supported. The small regions (a few light-days in size) inferred for several high ionization lines suggest that these lines arise from the accretion disk. In the FeIII$\lambda\lambda$2039-2113 blend, a  spectral feature relatively uncontaminated by other species, we measure very large microlensing variability (comparable in some extreme cases to that typical of X-ray) and the largest kinematic broadening. These results suggest  that FeIII (and likely other high ionization species present in strongly microlensed complexes) may arise in part from an inner region of the accretion disk. RM studies of the strongly microlensed iron spectral features could be of great interest in probing the accretion disk of quasars.\\ 

\acknowledgments
We thank the anonymous referee for valuable suggestions. C.F. gratefully acknowledges the financial support of a La Caixa PhD fellowship. E.M. is supported by the Spanish MINECO with the grants AYA2013-47744-C3-3-P and AYA2013-47744-C3-1-P. J.A.M. is supported by the Generalitat Valenciana with the grant PROMETEO/2014/60. J.J.V. is supported by the Spanish Ministerio de Econom\'\i a y Competitividad and the Fondo Europeo de Desarrollo Regional (FEDER) through grant AYA2014-53506-P and by the Junta de Andaluc\'\i a through project FQM-108.\\

\bibliographystyle{aasjournal}
\bibliography{bib_paper}
\appendix 
\section{Data Analysis Methods \label{AA}}

For each of the brightest emission lines (CIV, CIII], and MgII) we fit a straight line $y=a\lambda+b$ to the continuum on either side of the emission line and subtract it from the spectrum. For all images and all epochs we normalize the continuum-subtracted spectra to match the core of the emission line defined by the flux within a narrow interval ($\pm$6\AA) centered on the peak of the line. The magnitude differences of the wings are then constructed from the fluxes found after subtracting the linear model for the continuum emission underneath the line profile. We estimate the average wing emission in different wavelength intervals ($\sim$ 25\AA\ for CIV, $\sim$ 35\AA\ for CIII] and MgII) on either side of the emission line peak, corresponding to velocity intervals of $\sim$ 4500 km/s for CIV, $\sim$ 5300 km/s for CIII], and $\sim$ 3600 kms/s for MgII. We have separated the line core from the wings by a buffer of $\pm 9$\AA\ to prevent underestimation of the microlensing in the wings. In those cases in which the emission line is affected by absorption lines an integration window avoiding absorption features was chosen. See the core matched spectra corresponding to CIV, CIII], and MgII in Figures \ref{spectra1} to \ref{spectra4}. We use the following statistics to calculate the magnitude difference between two different images/epochs (x,y):
\begin{equation}
d_i = w_i * (y_i-x_i),
\end{equation}
with weights $w_i=\sqrt{<y_i+x_i>/(y_i+x_i)}$, selected to equalize the typical deviations of the differences. From the mean value in a given wavelength interval, $\langle d_i \rangle$, we compute the magnitude difference between images/epochs, $d = \langle d_i \rangle$, and its standard deviation $\sigma$ (see Tables \ref{1.5em_images}-\ref{1.5o_epochs}).\\

We have also analyzed the wavelength regions between CIV, CIII], and MgII to measure the changes in the UV FeII and FeIII emission line blends, the complex formed by the HeII line, the OIII]/AlII blend and the subjacent pseudo-continuum, and the red shelf of CIV. We follow the definition of the wavelength regions of \citealt{Guerras2013a,Guerras2013b}, \citealt{Vestergaard2001}, and \citealt{VandenBerk2001} and use the cores of the CIV, CIII], and MgII emission lines as a baseline for no microlensing. We fit various straight lines to the continuum regions bracketing the emission line windows and subtract them from the spectra. Then, for each image pair and each epoch, we normalize the continuum-subtracted spectra to match the core of the MgII (CIII]) emission line. In many cases, the MgII (CIII]) based normalization does not match the CIII] (CIV) emission line. We assume that this mismatch of the line cores arises from differential extinction in the lens galaxy. This is corrected by applying a linear extinction correction to match both emission lines simultaneously (obviously, this correction is applied to the data in \S 2.2. but not in \S 2.3.). Finally, for each pair of images (and each epoch) we compare the flux ratios in the defined emission line windows of the continuum-subtracted and extinction-corrected spectra using the same statistics as described before. See the resulting core matched spectra and chosen integration windows in Figures \ref{iron1} to \ref{iron5}.\\

\section{Intrinsic Variability and Microlensing in the CIV to MgII  Wavelength Region \label{AB}}

In the wavelength region between CIV and CIII], there is evidence of variability at the 2$\sigma$ level that very noticeably affects the complex formed by the HeII line, the OIII]/AlII blend, and the subjacent pseudo-continuum. We find microlensing in HE 0047-1756, HE 0435-1223, SDSS J1004+4112, and SDSS J1339+1310 and intrinsic variability in Q 0142-100, HE 0435-1223, and HE 2149-2745 (with partial evidence in QSO 0957+561 and HE 1104-1805). In the AlIII lines there is microlensing in SDSS J1004+4112 and partial evidence of intrinsic variability in HE 0435-1223. In the SiIII lines there is evidence of microlensing in HE 0435-1223, SDSS J1004+4112, HE 1104-1805, and WFI J2033-4723, and partial evidence of intrinsic variability in HE 0435-1223. Finally, the FeII blends included in this spectral range, show no evidence of intrinsic variability or microlensing.\\

{ In the wavelength region between CIII] and MgII there is also evidence of intrinsic variability and microlensing affecting several lines and complexes, particularly the FeII and FeIII iron lines. In FeII we detect (2$\sigma$ level) microlensing in HE 0435-1223, SDSS J0806+2006, FBQS J0951+2635 QSO 0957+561 (this can be also interpreted as intrinsic variability plus a time delay), and SDSS J1339+1310, and intrinsic variability in HE 0047-1756, HE 0435-1223, WFI J2033-4723 (partial evidence), and HE 2149-2745. In the FeIII$\lambda\lambda2040-2100$ blend, we obtain basically the same results, microlensing in HE 0435-1223, SDSS J0806+2006, FBQS J0951+2635, QSO 0957+561  (this can be also interpreted as intrinsic variability plus a time delay), and SDSS J1339+1310, and intrinsic variability in HE 0047-1756, HE 0435-1223, SDSS J1339+1310, WFI J2033-4723, and HE 2149-2745 (partial evidence).}

\begin{turnpage}
\begin{figure*}
\includegraphics[width=22cm]{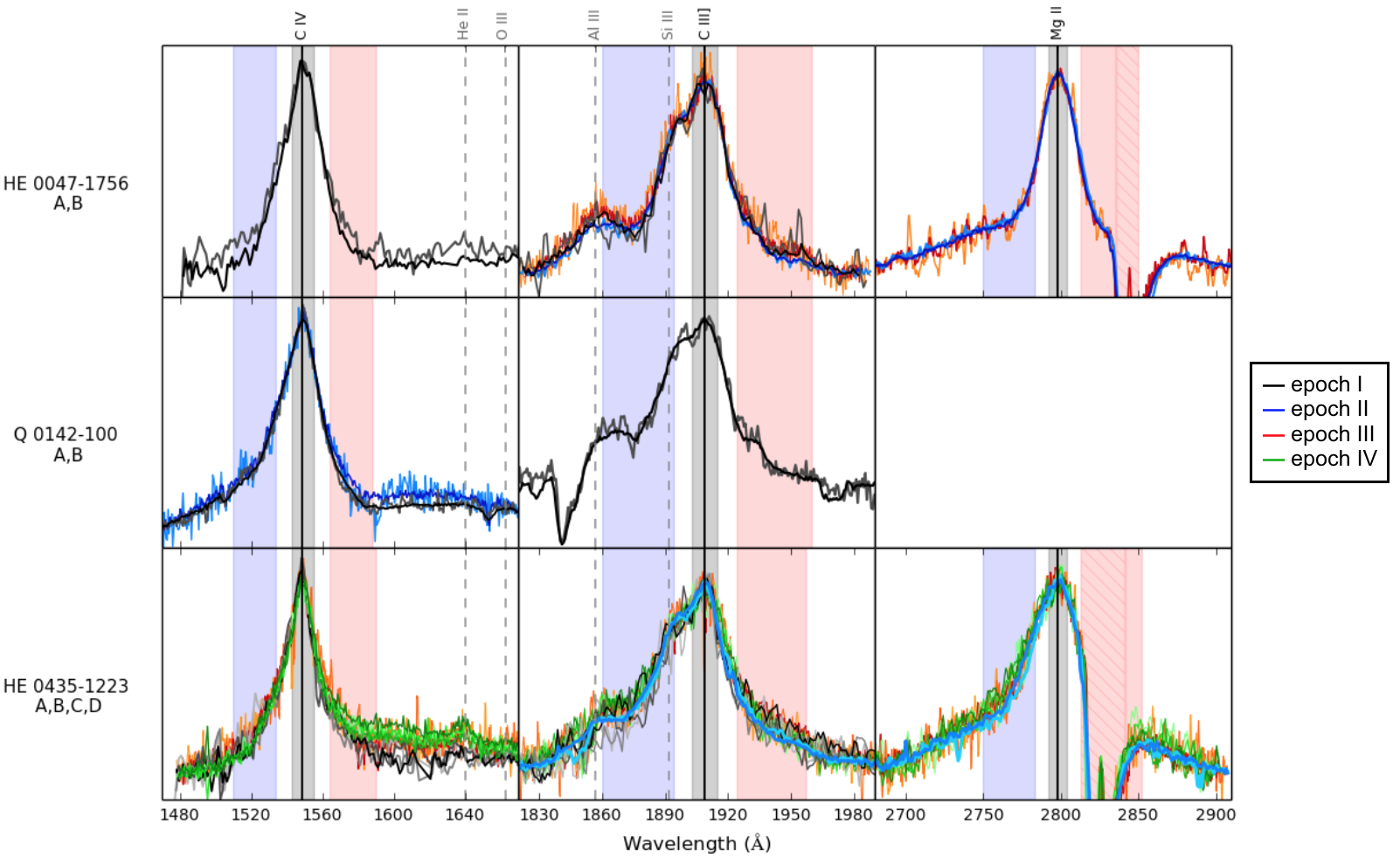}
\caption{CIV, CIII], and Mg II emission line profiles from different epochs superimposed after subtracting the continuum and matching the line core.
Different colors correspond to different epochs and different color shadings stand for different images in the corresponding epoch. Blue/red shaded regions show the integration windows used for the magnitude difference calculations. The ordinate is in arbitrary units of flux.}
\label{spectra1}
\end{figure*}
\end{turnpage}

\begin{turnpage}
\begin{figure*}
\includegraphics[width=22cm]{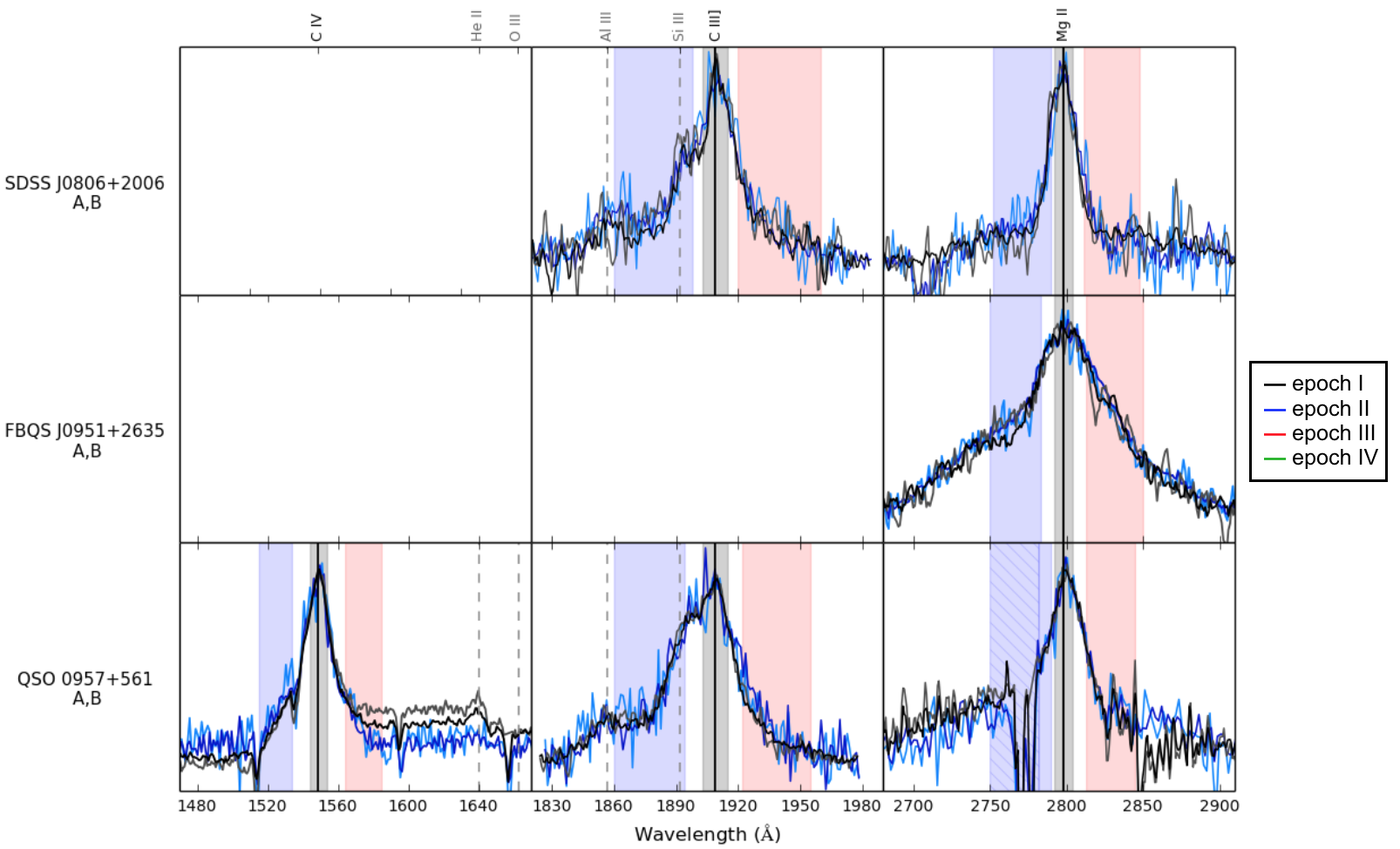}
\caption{Continuation of Figure \ref{spectra1}.}
\label{spectra2}
\end{figure*}
\end{turnpage}

\begin{turnpage}
\begin{figure*}
\includegraphics[width=22cm]{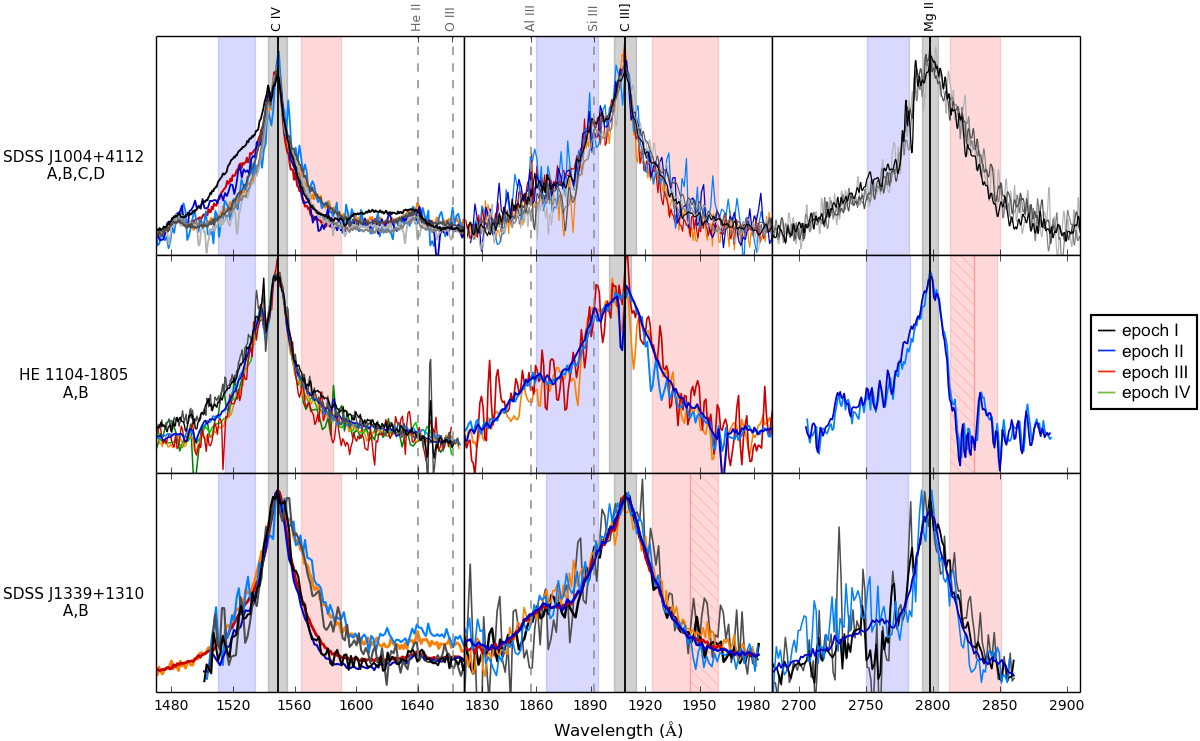}
\caption{Continuation of Figure \ref{spectra1}. Note: due to the lack of wavelength coverage towards the blue the definition of the continuum window on this side slightly overlaps with the blue wing in the case of SDSS J1339+1310 (epoch I and II).}
\label{spectra3}
\end{figure*}
\end{turnpage}

\begin{turnpage}
\begin{figure*}
\includegraphics[width=22cm]{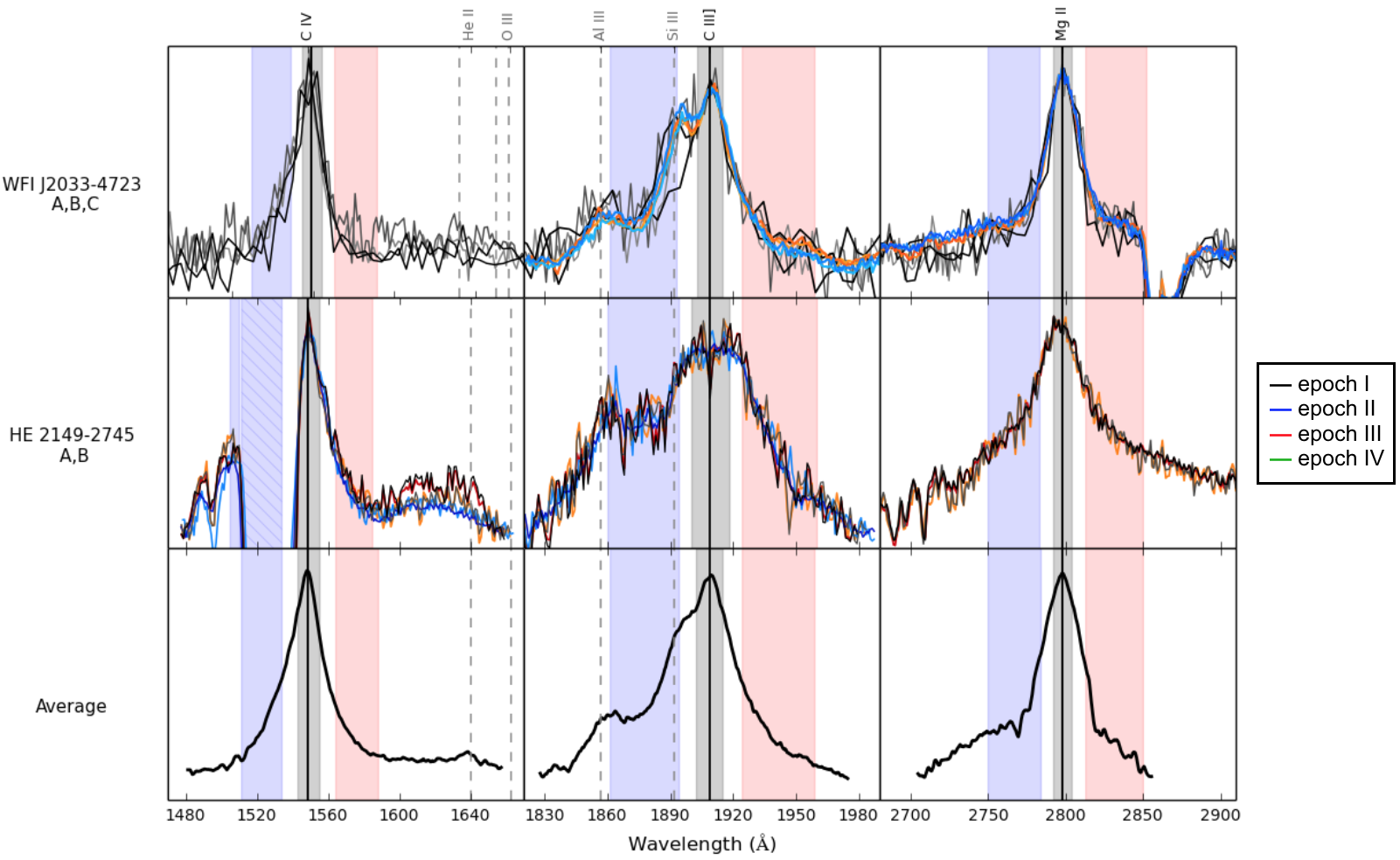}
\caption{Continuation of Figure \ref{spectra1}.}
\label{spectra4}
\end{figure*}
\end{turnpage}

\begin{turnpage}
\begin{figure*}
\centering
\includegraphics[width=21.5cm]{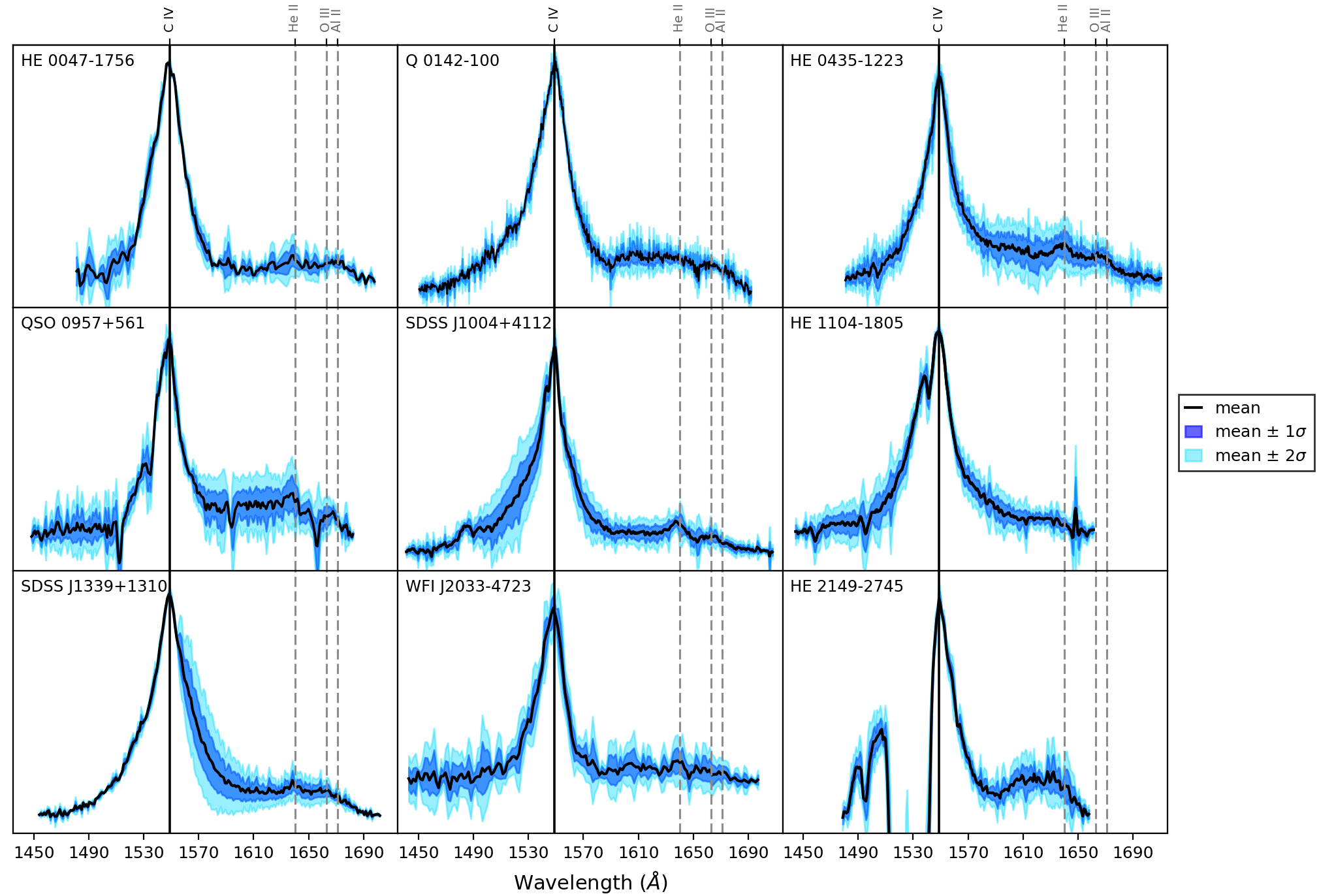}
\caption{Average (black) spectra for the CIV emission line with one (dark blue) and two (light blue) sigma intervals.}
\label{mean+sigma}
\end{figure*}
\end{turnpage}

\begin{figure*}
\centering
\includegraphics[width=19cm]{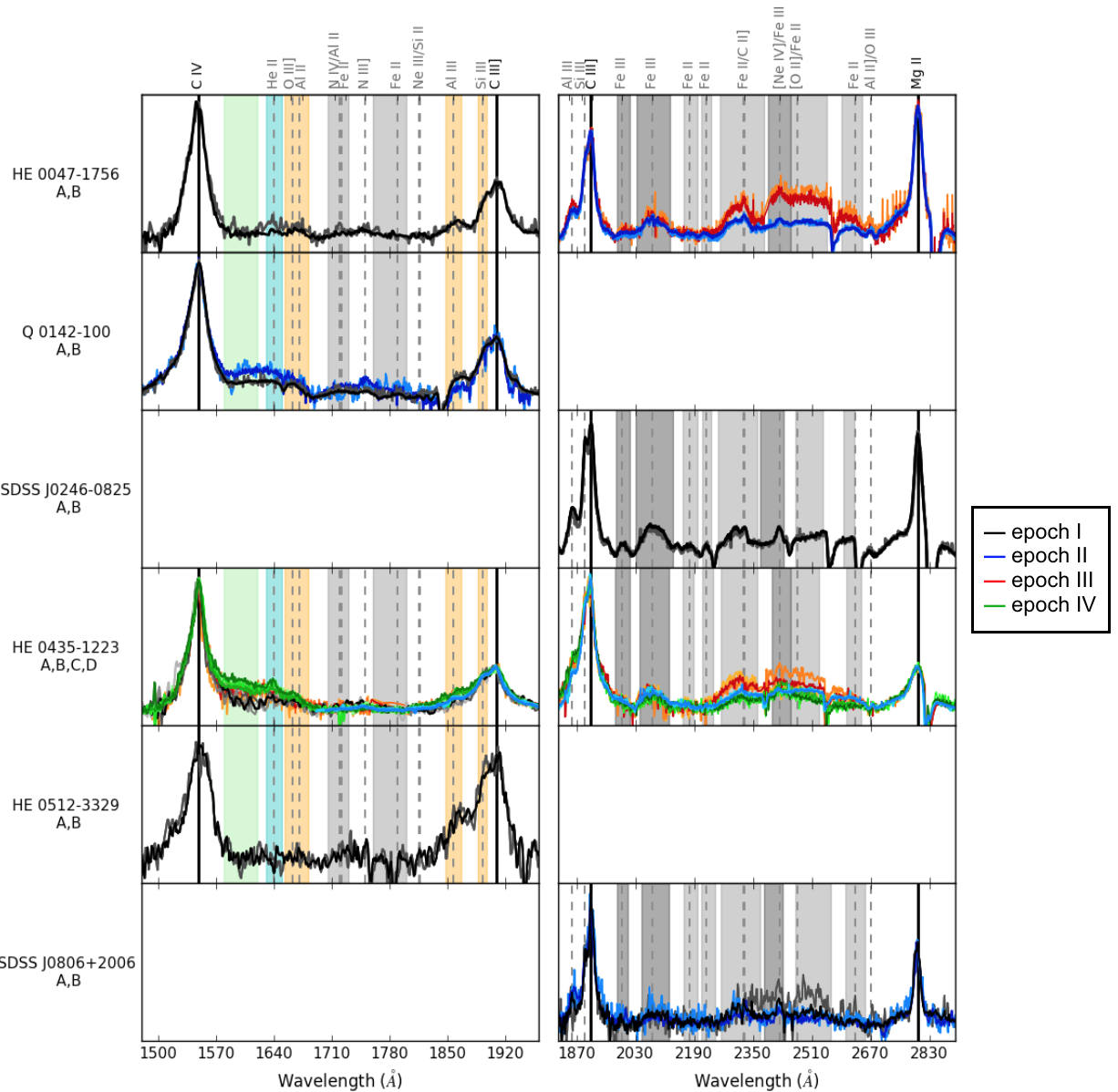}
\caption{Wavelength regions between CIV, CIII], and Mg II from different epochs superimposed after continuum subtraction, matching the line cores and extinction correction. Different colors correspond to different epochs and different color shadings stand for different images in the corresponding epoch. Shaded regions of different color show the integration windows used for the magnitude difference calculation of different emission line features. The ordinate is in arbitrary units of flux.}
\label{iron1}
\end{figure*} 

\begin{figure*}
\centering
\includegraphics[width=19cm]{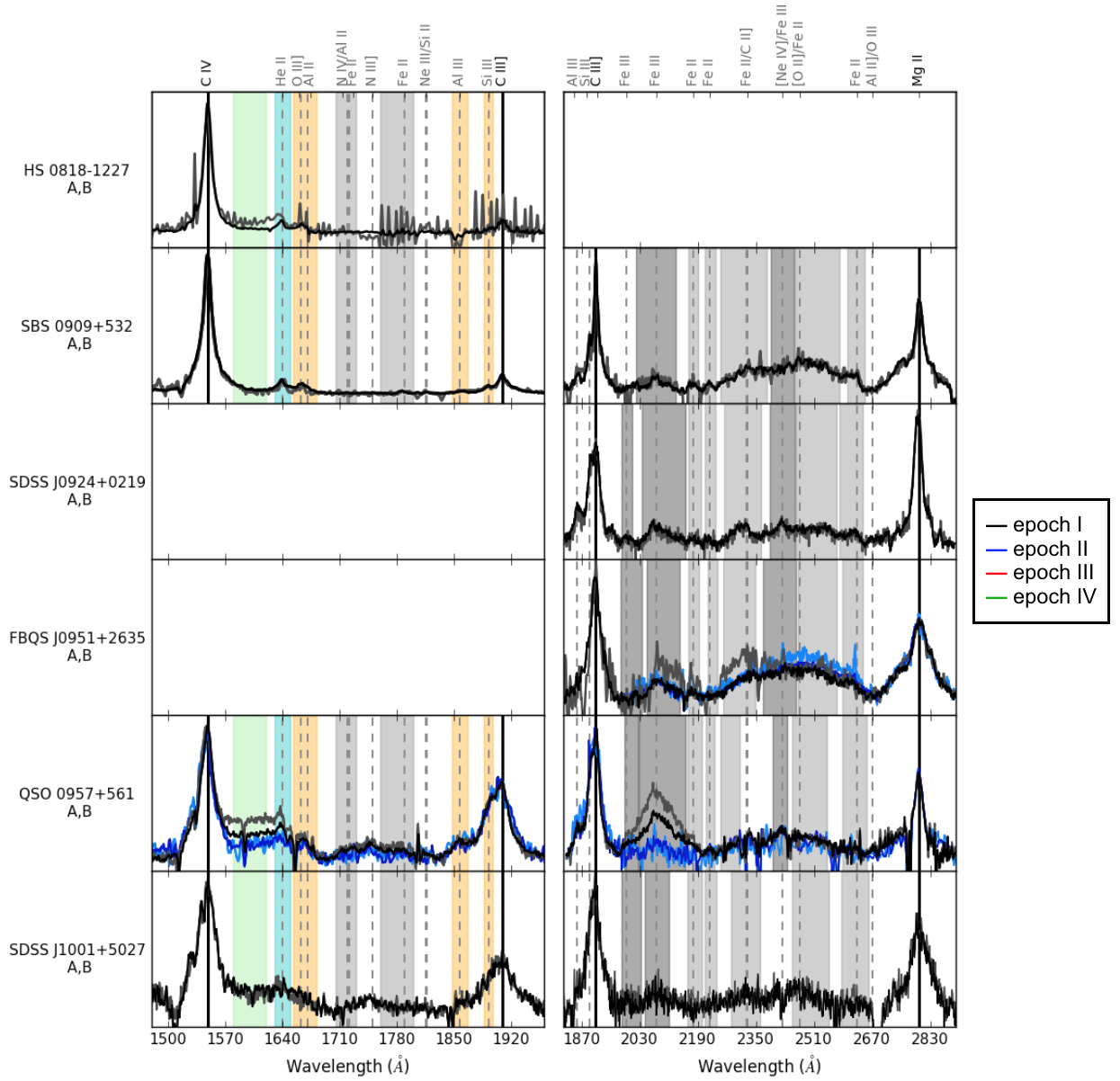}
\caption{Continuation of Figure \ref{iron1}.}
\label{iron2}
\end{figure*}

\begin{figure*}
\centering
\includegraphics[width=19cm]{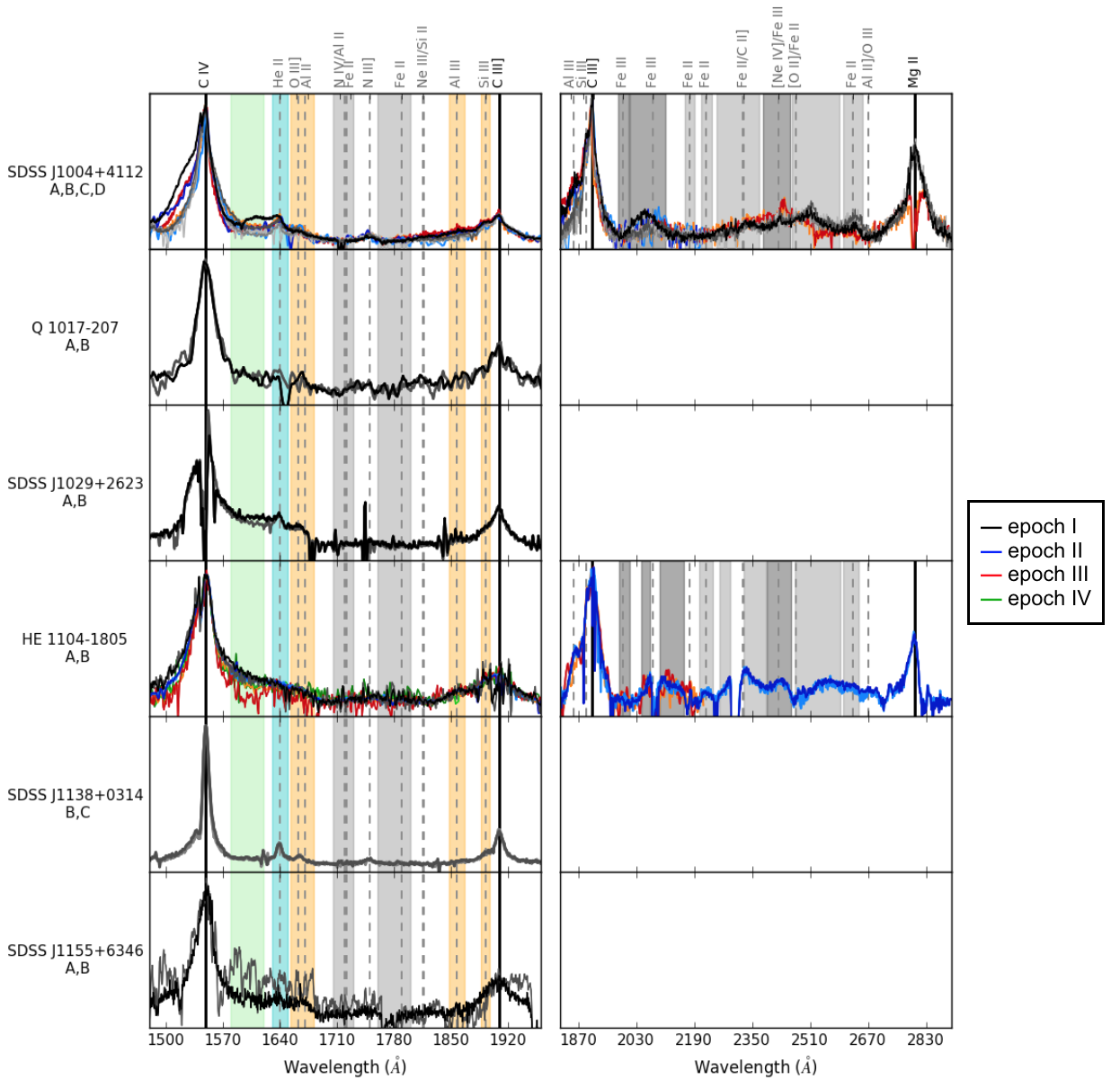}
\caption{Continuation of Figure \ref{iron1}.}
\label{iron3}
\end{figure*}

\begin{figure*}
\centering
\includegraphics[width=19cm]{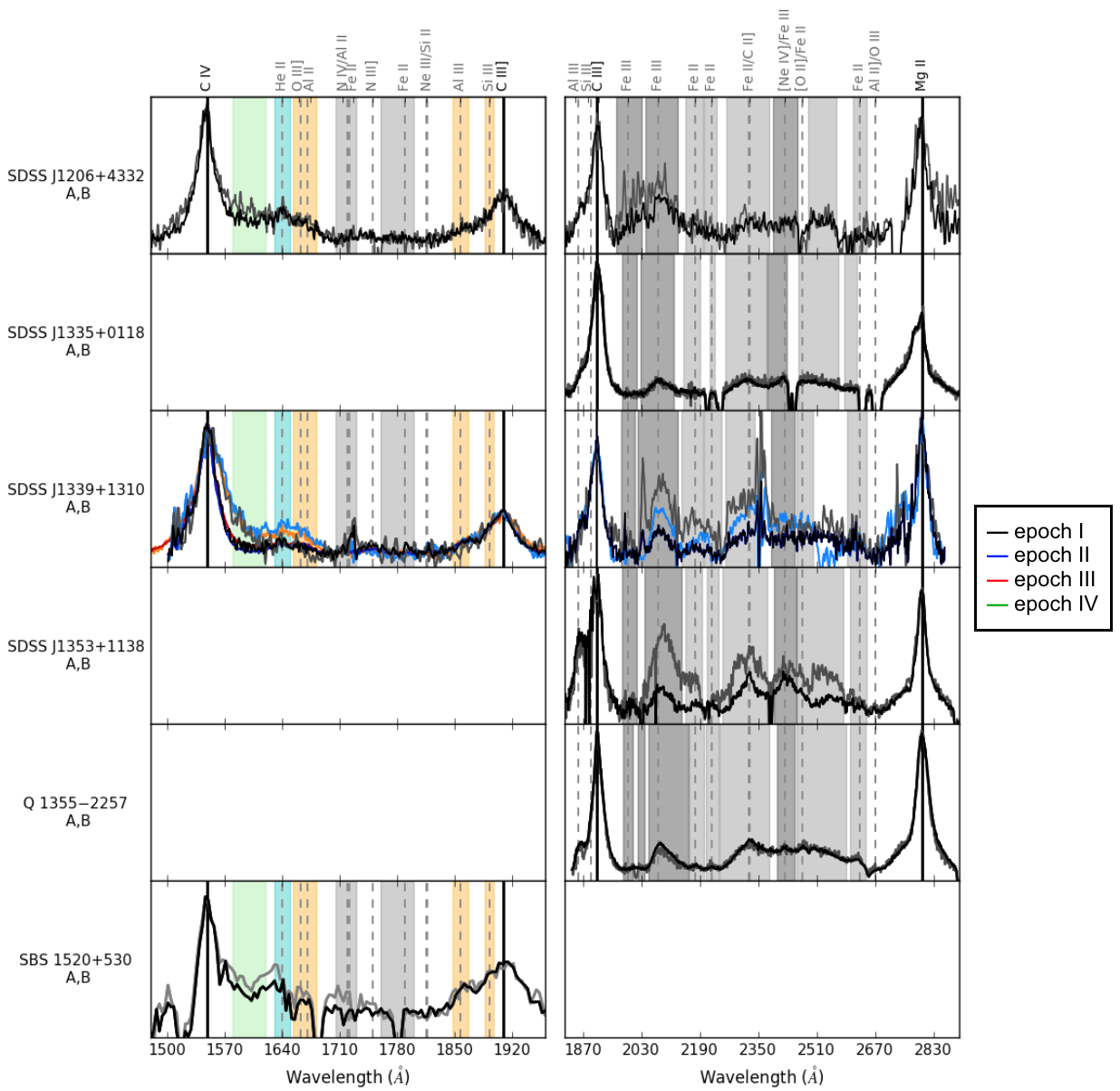}
\caption{Continuation of Figure \ref{iron1}.}
\label{iron4}
\end{figure*}

\begin{figure*}
\centering
\includegraphics[width=19cm]{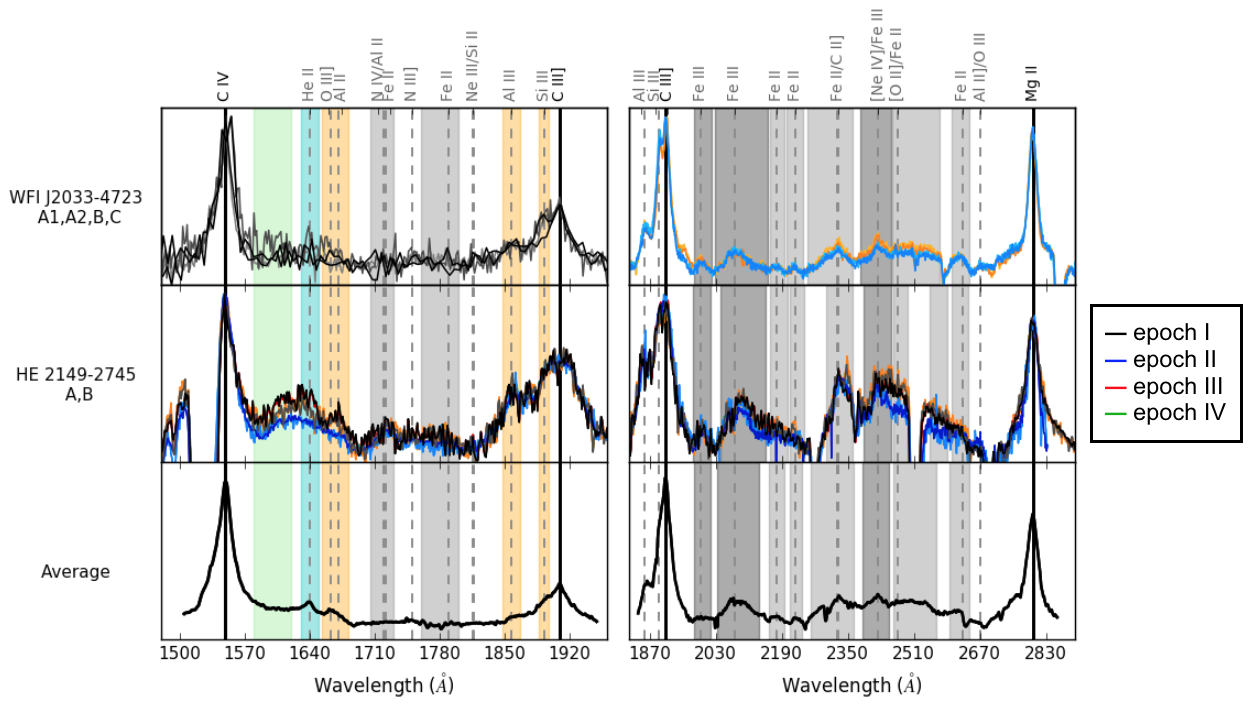}
\caption{Continuation of Figure \ref{iron1}.}
\label{iron5}
\end{figure*}

\begin{figure*}
\centering
\includegraphics[width=13.1cm]{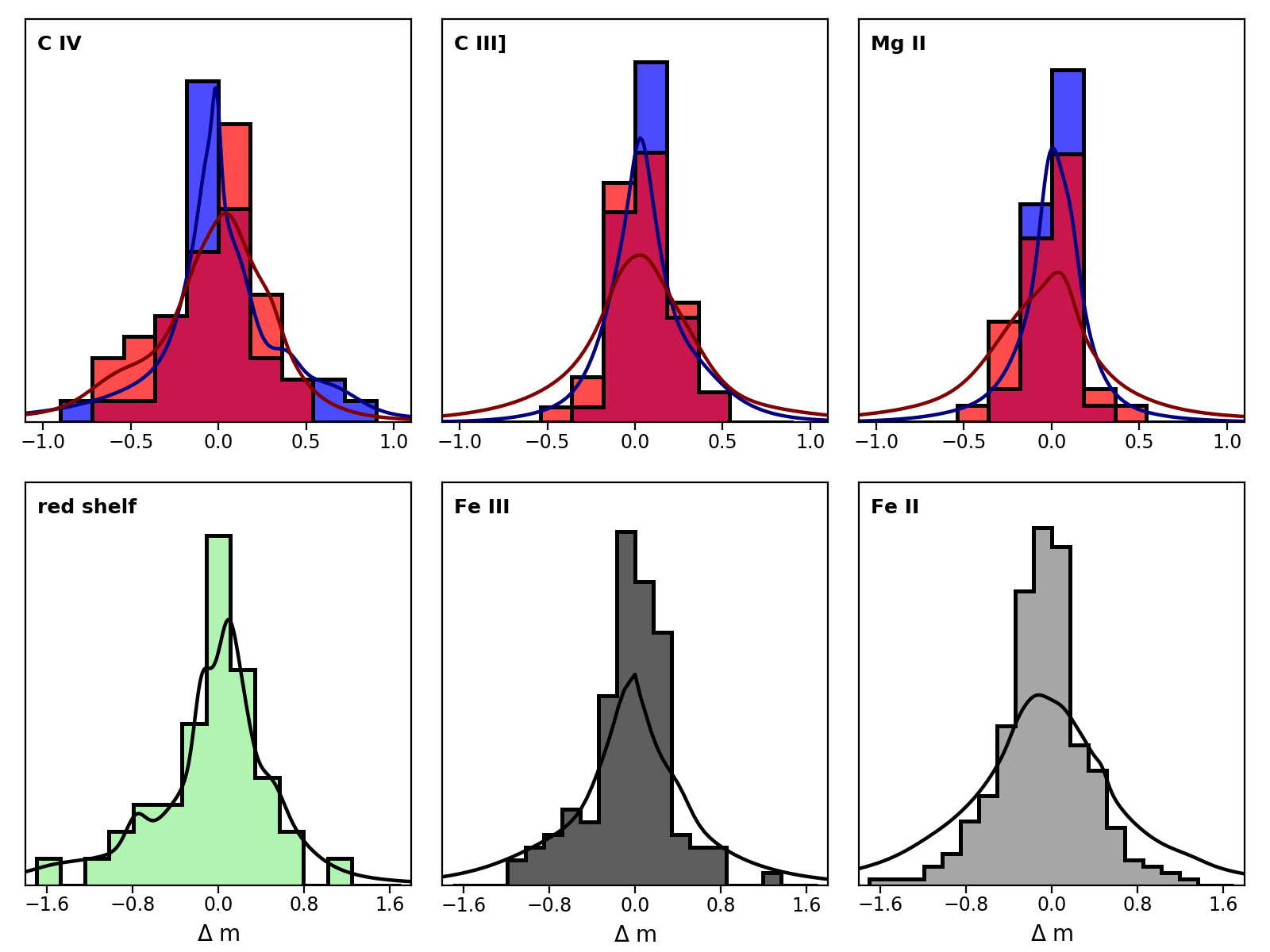}
\caption{Histograms of microlensing magnification difference between images for different emission line features. Black curves show the corresponding Gaussian kernel density estimates of the PDFs. For the broad emission lines CIV, CIII], and MgII (top row) histograms of the blue and red wings are shown. The blue (red) curve illustrates the Gaussian density estimate for the blue (red) wing.}
\label{histo_images}
\end{figure*}

\begin{figure*}
\centering
\includegraphics[width=13.1cm]{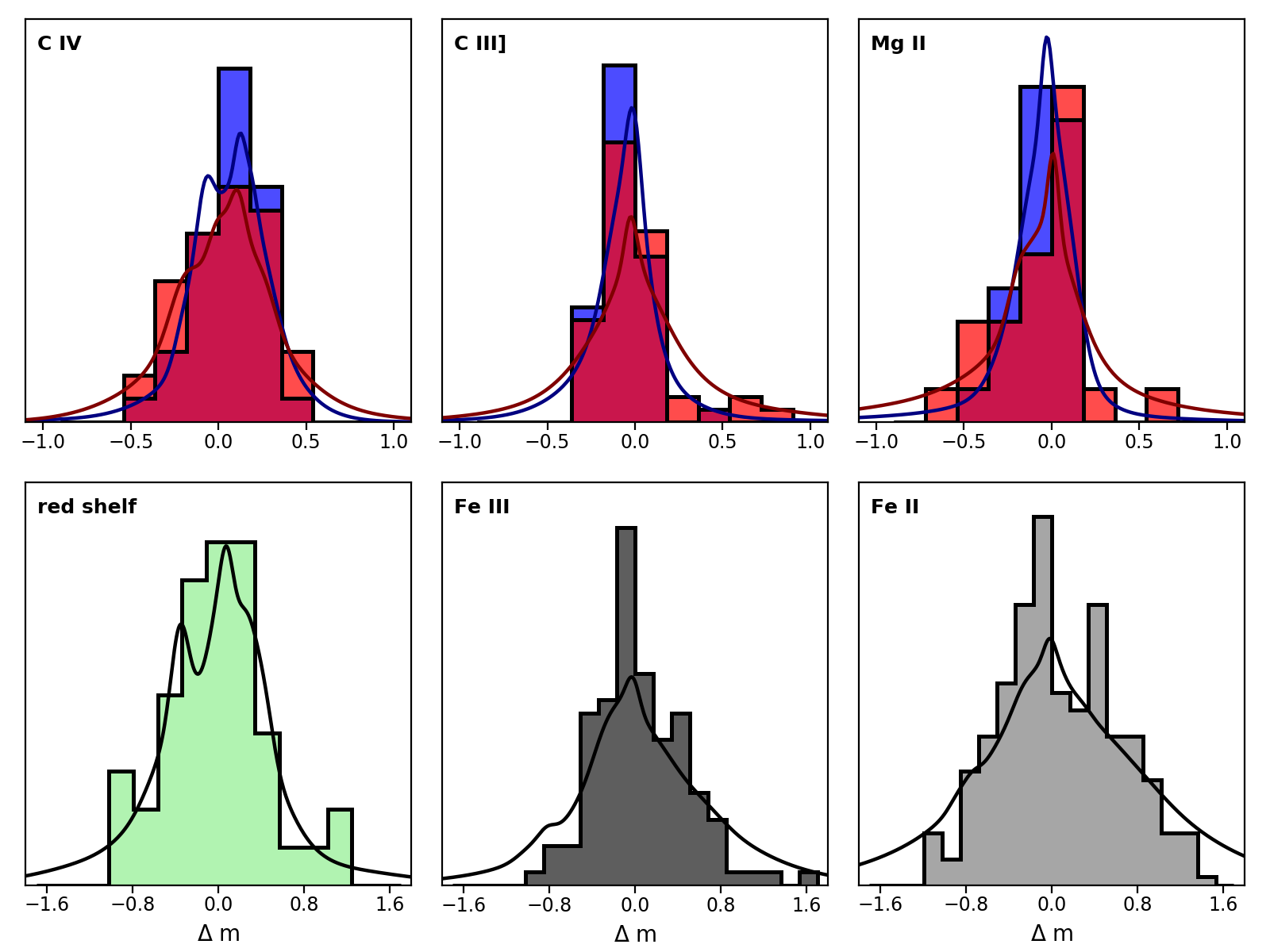}
\caption{Histograms of the difference between epochs for different emission line features. Black curves show the corresponding Gaussian kernel density estimates of the PDFs. For the broad emission lines CIV, CIII], and MgII (top row) histograms of the blue and red wings are shown. The blue (red) curve illustrates the Gaussian density estimate for the blue (red) wing.}
\label{histo_epochs}
\end{figure*}

\begin{figure*}
\centering
\includegraphics[width=17.5cm]{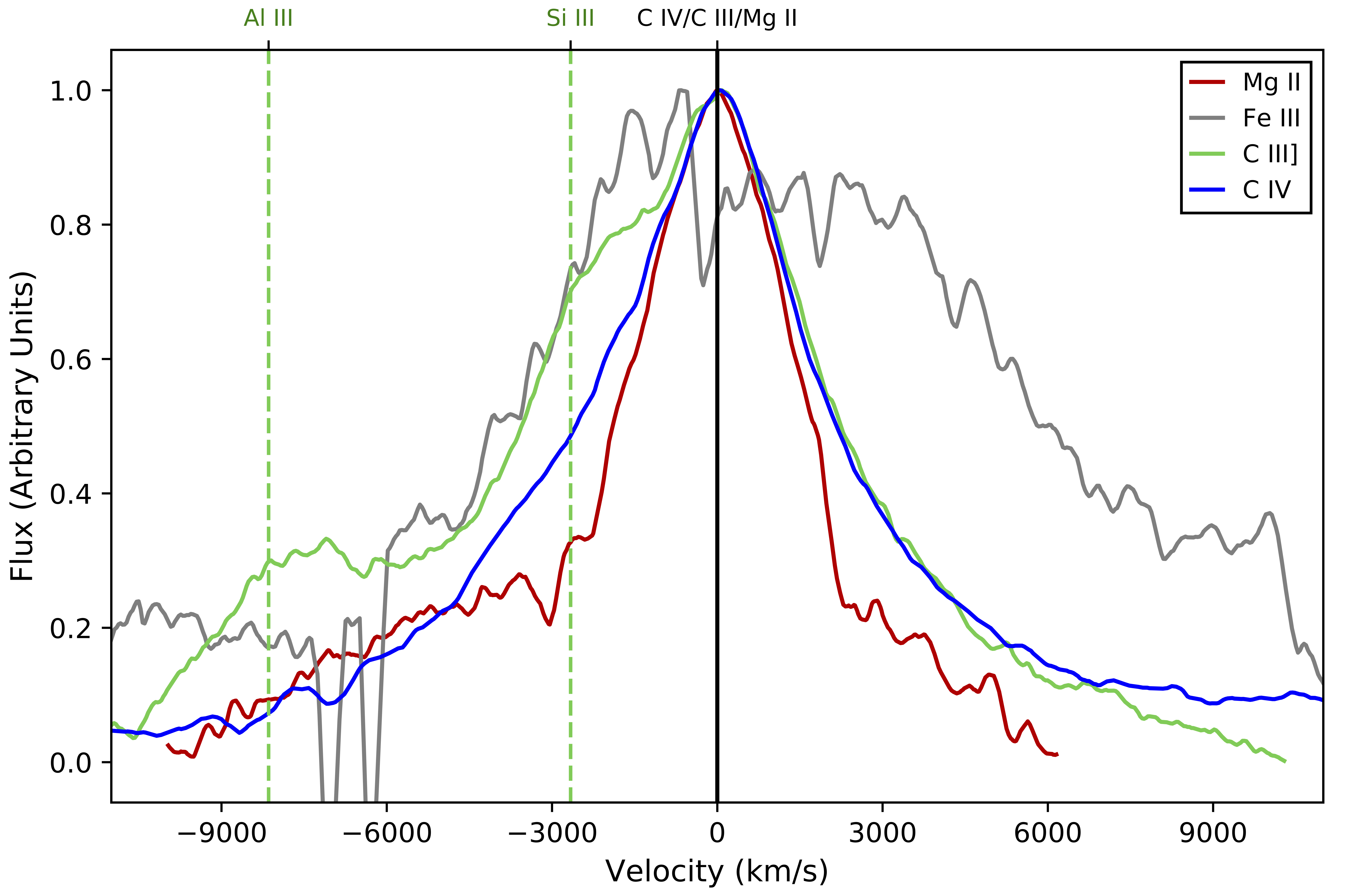}
\caption{Average line profiles of CIV, CIII], MgII, and FeII$\lambda\lambda$2040-2100 as a function of velocity.}
\label{line_velocity_matched}
\end{figure*}

\begin{figure*}[h]
\centering
\includegraphics[width=10cm]{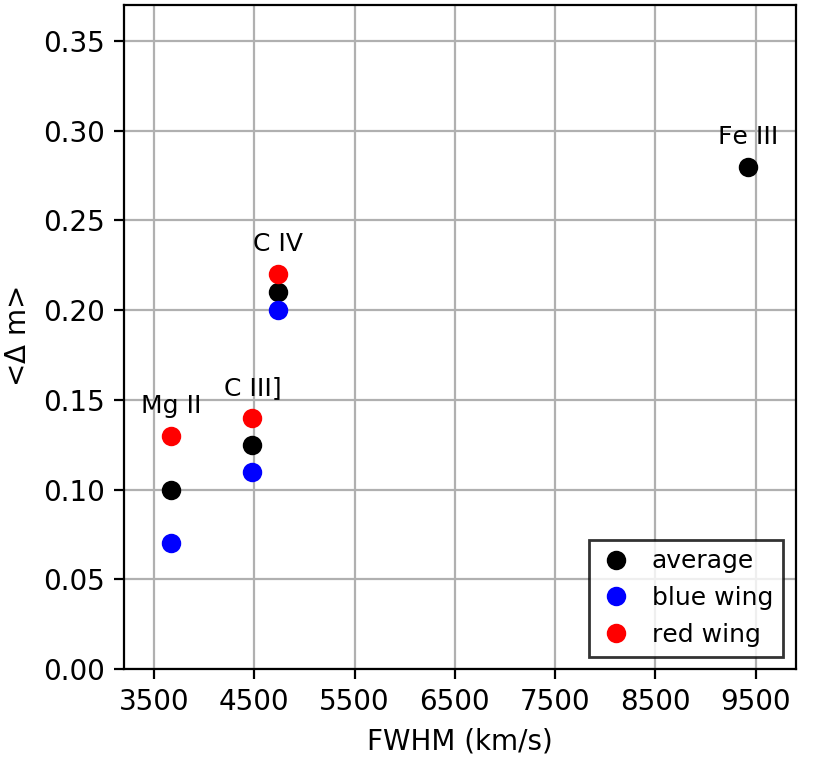}
\caption{Average amplitude of microlensing between images as a function of the line broadening for CIV, CIII], MgII, and FeIII.}
\label{m_vs_fwhm}
\end{figure*}

\begin{figure*}[h]
\centering
\includegraphics[width=14cm]{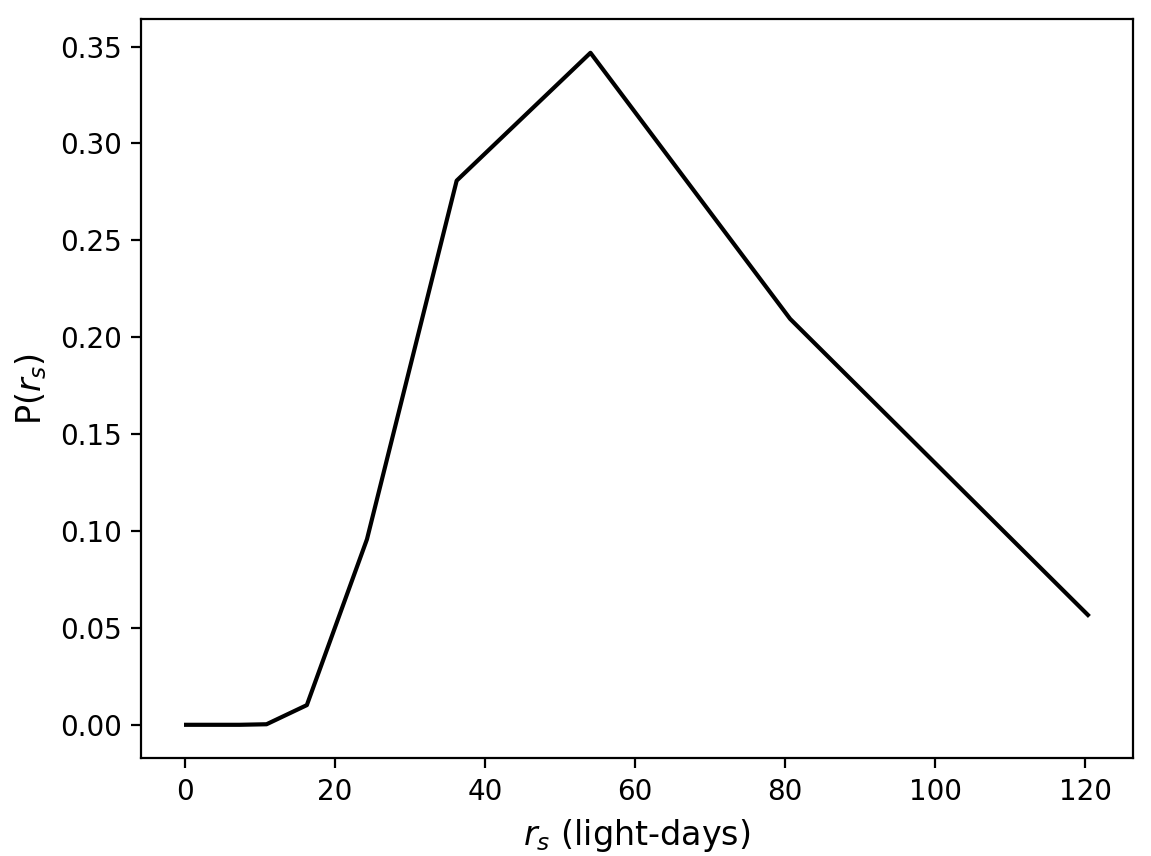}
\caption{Joint Likelihood for the low ionization lines CIII] and MgII.}
\label{PDF_lines}
\end{figure*}

\begin{figure*}[h]
\centering
\includegraphics[width=14cm]{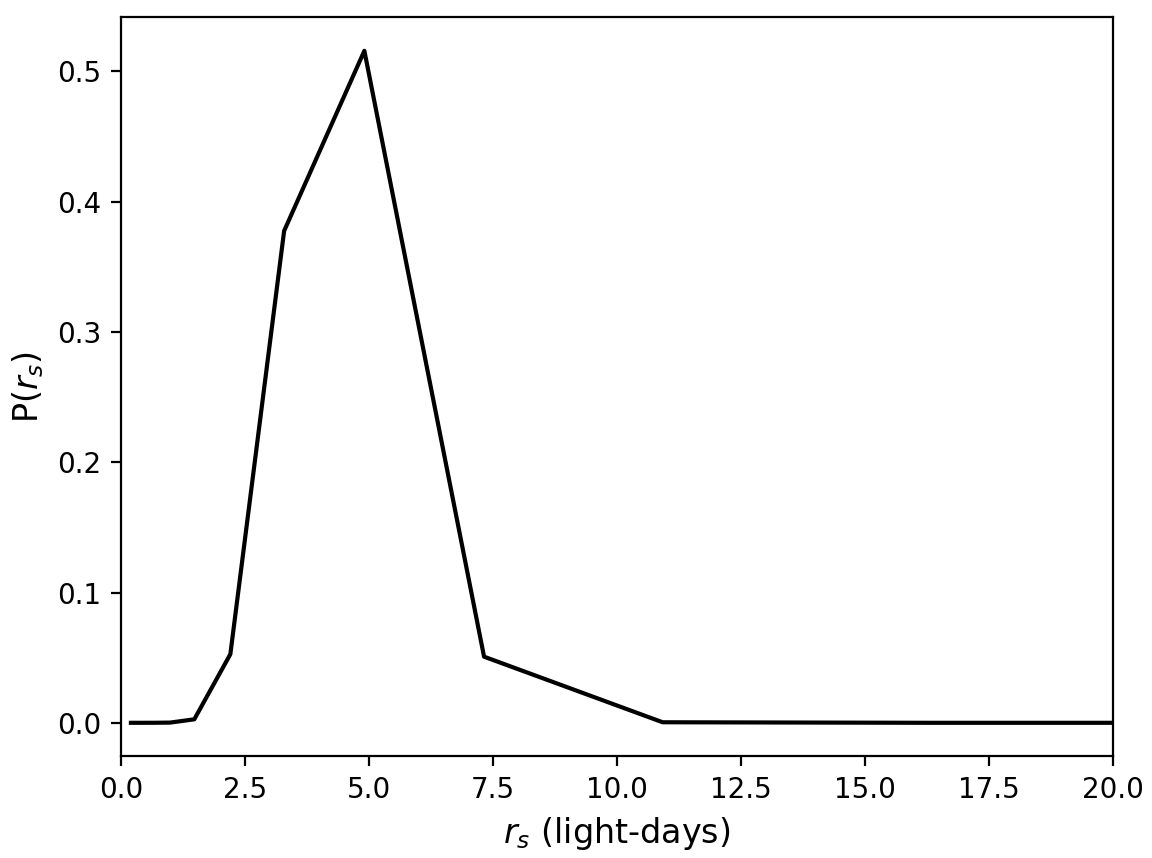}
\caption{Joint Likelihood for the red shelf.}
\label{PDF_other}
\end{figure*}

\begin{figure*}[h]
\centering
\includegraphics[width=13.7cm]{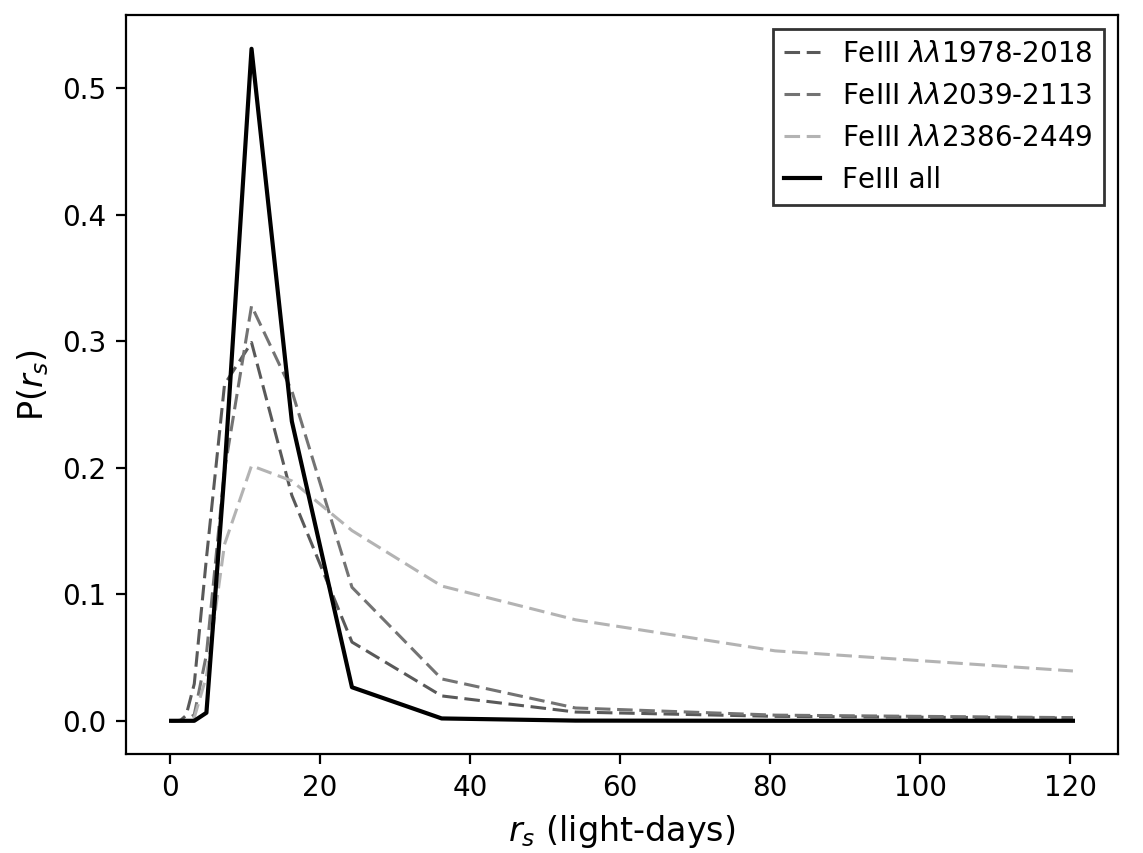}
\caption{Joint Likelihood for FeIII (solid black line). Dashed lines show the Joint Likelihood functions for different FeIII emitting regions (see Fig. \ref{iron1}-\ref{iron5}).}
\label{PDF_fe3}
\end{figure*}

\begin{figure*}[h]
\centering
\includegraphics[width=13.7cm]{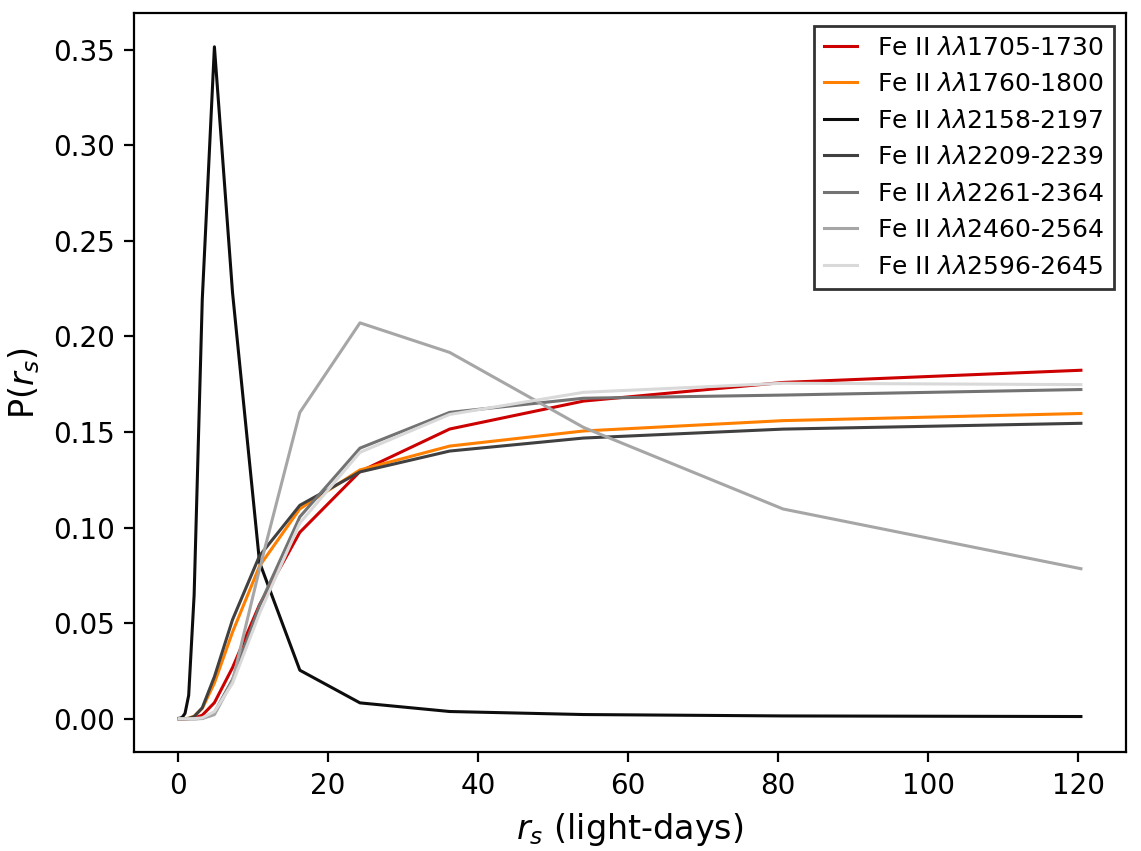}
\caption{Joint Likelihood functions for different FeII emitting regions (see Fig. \ref{iron1}-\ref{iron5}).}
\label{PDF_fe2}
\end{figure*}

\begin{deluxetable*}{lcccccccccc}
\tablewidth{\textwidth}
\setlength{\tabcolsep}{3.pt}
\renewcommand{\arraystretch}{0.8}
\tablecaption{Database of Lensed Quasar Spectra}
\tablehead{
\colhead{Object} & \colhead{Image} & \colhead{Epoch} & \colhead{Date} & \colhead{C IV} & \colhead{C III]} & \colhead{Mg II} & \colhead{Region 1$^{*}$} & \colhead{Region 2$^{**}$} & \colhead{Facilities} & \colhead{Reference}}
\startdata
	\multirow{3}{*}{HE 0047-1756} & \multirow{3}{*}{A,B} & I & 2002 Sep 04 & x & x & - & x & -  & Magellan & \citealt{Wisotzki2004}\\
	& & II & 2005 Jul 18 & - & x & x & - & x & VLT & \citealt{Sluse2012}\\
    & & III & 2008 Jan 13 & - & x & x & - & x & Magellan & \citealt{Rojas2014}\\ \hline \vspace*{-3mm}\\	
	\multirow{2}{*}{Q 0142-100} & \multirow{2}{*}{A,B} & I & 2006 Aug 15 & x & x & - & x & - & VLT & \citealt{Sluse2012}\\
	& & II & 2008 Jan 12 & x & - & - & x & x & MMT & Motta (priv. comm.)\\ \hline \vspace*{-3mm}\\ 	
    SDSS J0246-0825 & A,B & I & 2006 Aug 22 & - & - & - & - & x & VLT & \citealt{Sluse2012}\\ \hline \vspace*{-3mm}\\	
	\multirow{4}{*}{HE 0435-1223} & A,B & I & 2002 Sep 05 & x & x & x & x & - & CAO & \citealt{Wisotzki2003}\\ \cline{2-9} \vspace{-3mm}\\
	& B,D & II & 2004 Oct-Nov & - & x & x & x & x & VLT & \citealt{Eigenbrod2007}\\ \cline{2-9} \vspace{-3mm}\\
	& A,B,C,D & III & 2008 Jan 12 & x & x & x & x & x & MMT & \citealt{Motta2017}\\ \cline{2-9} \vspace*{-3mm}\\	
    & A,C,D & IV & 2007 Dec 10 & x & x & - & x & x & Magellan & Motta (priv. comm.)\\ \hline \vspace{-3mm}\\
     HE 0512-3329 & A,B & I & 2001 Aug 13 & - & - & - & x & - & HST & \citealt{Wucknitz2003}\\ \hline \vspace*{-3mm}\\	
	\multirow{2}{*}{SDSS J0806+2006} & \multirow{2}{*}{A,B} & I & 2005 Apr 12 & - & x & x & - & x & APO & \citealt{Inada2006}\\
	& & II & 2006 Apr 22 & - & x & x & - & x & VLT & \citealt{Sluse2012}\\ \hline \vspace*{-3mm}\\	
	HS 0818-1227 & A,B & I & 2008 Jan 12 & - & - & - & x & - & MMT & \citealt{Motta2012} \\ \hline \vspace*{-3mm}\\
	SBS 0909+532 & A,B & I & 2003 Mar 07 & - & - & - & x & x & HST & \citealt{Mediavilla2005}\\ \hline \vspace*{-3mm}\\ 	
	SDSS J0924+0219 & A,B & I & 2005 Jan 14 & - & - & - & - & x & VLT & \citealt{Eigenbrod2006}\\ \hline \vspace*{-3mm}\\	
    \multirow{2}{*}{FBQ 0951+2635} & \multirow{2}{*}{A,B} & I & 1997 Feb 14 & - & - & x & - & x & Keck & \citealt{Schechter1998}\\
	& & II & 2006 Mar 31 & - & - & x & - & - & VLT & \citealt{Sluse2012}\\ \hline \vspace*{-3mm}\\	
	\multirow{3}{*}{Q 0957+561} & A & I & 1999 Apr 15 & x & x & x & x & x & HST & \citealt{Goicoechea2005}\\
	& B & I & 2000 Jun 2 & x & x & x & x & x & HST & \citealt{Goicoechea2005}\\
	& A,B & II & 2008 Jan 12 & x & x & x & x & - & MMT & \citealt{Motta2012} \\ \hline \vspace*{-3mm}\\	
    SDSS J1001+5027 & A,B & I & 2003 Nov 20 & - & - & - & x & x & APO & \citealt{Oguri2005}\\ \hline \vspace*{-3mm}\\
	\multirow{3}{*}{SDSS 1004+4112} & A,B,C,D & I & 2003 May 31 & x & x & x & x & x & APO & \citealt{Richards2004}\\ \cline{2-9} \vspace{-3mm}\\
	& \multirow{2}{*}{A,B} & II & 2004 Jan 19 & x & x & - & x & - & WHT & \citealt{Gomez2006}\\ 
	& & III & 2008 Jan 12 & x & x & - & x & x & MMT & \citealt{Motta2012}\\ \hline \vspace*{-3mm}\\	
    Q 1017-207& A,B & I & 1996 Oct 28 & - & - & - & x & - & HST & \citealt{Surdej1997}\\ \hline \vspace*{-3mm}\\
	SDSS J1029+2623 & A,B & I & 2008 Jan 12 & - & - & - & x & - & MMT & \citealt{Motta2012} \\ \hline \vspace*{-3mm}\\	
	\multirow{4}{*}{HE 1104-1805} & \multirow{4}{*}{A,B} & I & 2008 Jan 11 & x & - & - & x & - & MMT & \citealt{Motta2012}\\
	& & II & 2008 Apr 07 & x & x & x & x & x & VLT & \citealt{Motta2012} \\
	& & III & 1993 May 11 & x & x & - & x & - & NTT & \citealt{Wisotzki1993} \\
	& & IV & 1994 Nov 29 & x & - & - & x & - & ESO 3.6m & \citealt{Wisotzki1995} \\ \hline \vspace*{-3mm}\\	
    SDSS J1138+0314 & B,C & I & 2005 May 10 & - & - & - & x & - & VLT & \citealt{Sluse2012}\\ \hline \vspace*{-3mm}\\
	SDSS J1155+6346 & A,B & I & 2010 Sept 20 & - & - & - & x & - & HST & \citealt{Rojas2014} \\ \hline \vspace*{-3mm}\\
	SDSS J1206+4332 & A,B & I & 2004 Jun 21 & - & - & - & x & x & APO & \citealt{Oguri2005}\\ \hline \vspace*{-3mm}\\
	SDSS J1335+0118 & A,B & I & 2005 Feb 17 & - & - & - & - & x & VLT & \citealt{Sluse2012}\\ \hline \vspace*{-3mm}\\
	\multirow{3}{*}{SDSS J1339+1310} & \multirow{3}{*}{A,B} & I & 2013 Apr 13 & - & x & x & x & x & GTC & \citealt{Shalyapin2014}\\
	& & II & 2014 Mar 27 & - & x & x & x & x & GTC & \citealt{Goicoechea2016}\\
	& & III & 2014 May 20 & x & x & - & - & - & GTC & \citealt{Goicoechea2016}\\ \hline \vspace*{-3mm}\\	
    SDSS J1353+1138 & A,B & I & 2005 Apr 12 & - & - & - & - & x & Keck & \citealt{Inada2006}\\ \hline \vspace*{-3mm}\\
    Q 1355-2257 & A,B & I & 2005 Mar 13 & - & - & - & - & x & VLT & \citealt{Sluse2012}\\ \hline \vspace*{-3mm}\\
	SBS 1520+530 & A,B & I & 1996 Jun 12 & - & - & - & x & - & SAO & \citealt{Chavushyan1997}\\ \hline \vspace*{-3mm}\\
	\multirow{3}{*}{WFI 2033-4723} & A1,A2,B,C & I & 2003 Sep 15 & x & x & x & x & - & Magellan & \citealt{Morgan2004}\\ \cline{2-9} \vspace{-3mm}\\
	& \multirow{2}{*}{B,C} & II & 2005 May 13 & - & x & x & - & x & VLT & \citealt{Sluse2012}\\ 
	& & III & 2008 Apr 14 & - & x & x & - & x & VLT & \citealt{Motta2017} \\ \hline \vspace*{-3mm}\\	
	\multirow{3}{*}{HE 2149-2745} & \multirow{3}{*}{A,B} & I & 2000 Nov 19 & x & x & x & x & x & VLT & \citealt{Burud2002}\\
	& & II & 2006 Aug 04 & x & x & - & x & x & VLT & \citealt{Sluse2012}\\
	& & III & 2008 May 07 & x & x & x & x & x & VLT & \citealt{Motta2017}\\	
\enddata
\tablenotetext{*}{Region between C IV and C III]}
\tablenotetext{**}{Region between C III] and Mg II}
\end{deluxetable*}
\label{1}

\begin{deluxetable*}{lccccccc}
\tablewidth{\textwidth}
\renewcommand{\arraystretch}{0.704}
\tablecaption{Differences between Images - CIV, CIII] and MgII Lines}
\tablehead{\colhead{Emission Line}& \colhead{Wing}&\colhead{Object}&\colhead{Image Pair}&\colhead{Epoch}&\colhead{d$^{*}$ $\pm$ $\sigma^{**}$}& \colhead{d/$\sigma$}&\colhead{Classification}\vspace*{-1mm}} 
\startdata
	\multirow{14}{*}{C IV} & \multirow{5}{*}{blue wing} & \multirow{5}{*}{SDSS 1004+4112} & \multirow{2}{*}{B-A} & I & 0.56 $\pm$ 0.16 & 3.4 & Microlensing \\
	& & & & III & 0.39 $\pm$ 0.07 & 5.8 & Microlensing \\ \cline{4-8} \vspace*{-3mm}\\
	& & & C-A & I & 0.69 $\pm$ 0.14 & 4.8 & Microlensing \\ \cline{4-8} \vspace*{-3mm}\\
	& & & D-A & I & 0.90 $\pm$ 0.36 & 2.5 & Microlensing \\ \cline{4-8} \vspace*{-3mm}\\
	& & & C-B & I & 0.13 $\pm$ 0.06 & 2.2 & Microlensing \\ \cline{2-8} \vspace*{-3mm}\\
	& \multirow{9}{*}{red wing} & \multirow{3}{*}{HE 0435-1223} & B-A & IV & 0.20 $\pm$  0.07 & 2.8 & Microlensing \\ \cline{4-8} \vspace*{-3mm}\\
	& & & D-B & IV & -0.12 $\pm$ 0.08 & 1.5 & \\ \cline{4-8} \vspace*{-3mm}\\
	& & & D-C & I & -0.39 $\pm$ 0.27 & 1.5 & \\ \cline{3-8} \vspace*{-3mm}\\
	& & \multirow{5}{*}{SDSS 1004+4112} & \multirow{3}{*}{B-A} & I & 0.30 $\pm$ 0.06 & 5.1 & Microlensing \\
	& & & & II & -0.53 $\pm$ 0.31 & 1.7 & \\
	& & & & III & -0.52 $\pm$ 0.15 & 3.6 & Microlensing \\ \cline{4-8} \vspace*{-3mm}\\
	& & & C-B & I & 0.35 $\pm$ 0.10 & 3.5 & Microlensing \\ \cline{4-8} \vspace*{-3mm}\\
	& & & D-B & I & 0.40 $\pm$ 0.26 & 1.6 & \\ \cline{3-8} \vspace*{-3mm}\\
	& & SDSS J1339+1310 & B-A & III & -0.62 $\pm$ 0.15 & 4.2 & Microlensing \\
	\hline \vspace*{-3mm}\\
	\multirow{5}{*}{C III]} & \multirow{2}{*}{blue wing} & SDSS 1004+4112 & C-A & I & 0.33 $\pm$ 0.18 & 1.9 & \\ \cline{3-8} \vspace*{-3mm}\\
	& & WFI 2033-4723 & C-B & II & 0.12 $\pm$ 0.06 & 2.1 & Microlensing \\ \cline{2-8} \vspace*{-3mm}\\
	& \multirow{3}{*}{red wing} & \multirow{2}{*}{HE 0435-1223} & B-A & I & 0.29 $\pm$  0.14 & 2.0 & Microlensing \\ \cline{4-8} \vspace*{-3mm}\\
	& & & C-A & I & 0.29 $\pm$ 0.15 & 1.9 & \\ \cline{3-8} \vspace*{-3mm}\\
	& & HE 1104-1805 & B-A & II & 0.21 $\pm$  0.14 & 1.5 & \\
	\hline \vspace*{-3mm}\\
	\multirow{2}{*}{Mg II} & blue wing &  HE 0435-1223 & D-B & II & 0.11 $\pm$  0.04 & 3.0 & Microlensing \\ \cline{2-8} \vspace*{-3mm}\\
	& red wing & HE 0435-1223 & D-B & II & -0.27 $\pm$ 0.16 & 1.7 & \\
\enddata
\tablenotetext{*}{\small magnitude difference}
\tablenotetext{**}{\small standard deviation of magnitude difference}
\end{deluxetable*}
\label{1.5em_images}

\begin{deluxetable*}{lccccccc}
\tablewidth{\textwidth}
\renewcommand{\arraystretch}{0.704}
\tablecaption{Differences between Epochs - CIV, CIII] and MgII Lines}
\tablehead{\colhead{Emission Line}& \colhead{Wing}&\colhead{Object}&\colhead{Image}&\colhead{Epoch}&\colhead{d$^{*}$ $\pm$ $\sigma^{**}$}& \colhead{d/$\sigma$}&\colhead{Classification}\vspace*{-1mm}} 
\startdata
	\multirow{20}{*}{C IV} & \multirow{12}{*}{blue wing} & Q 0142-100 & A & II-I & -0.07 $\pm$ 0.05 & 1.5 & \\  \cline{3-8} \vspace*{-3mm}\\
	& & \multirow{3}{*}{SDSS 1004+4112} & \multirow{3}{*}{A} & II-I & 0.28 $\pm$ 0.07 & 4.1 & Microlensing Variability\\
	& & & & III-I & 0.18 $\pm$ 0.04 & 4.2 & Microlensing Variability\\
	& & & & III-II & -0.10 $\pm$ 0.06 & 1.5 & \\  \cline{3-8} \vspace*{-3mm}\\
	& & \multirow{8}{*}{HE 1104-1805} &\multirow{4}{*}{ A} & II-I & 0.14 $\pm$ 0.08 & 1.7 & \\
	& & & & III-I & 0.26 $\pm$ 0.14 & 1.9 & \\
	& & & & IV-I & 0.28 $\pm$ 0.15 & 1.9 & \\
	& & & & IV-II & 0.19 $\pm$ 0.10 & 1.9 & \\ \cline{4-8} \vspace*{-3mm}\\
	& & & \multirow{4}{*}{B} & II-I & 0.18 $\pm$ 0.11 & 1.6 & \\
	& & & & III-I & 0.36 $\pm$ 0.21 & 1.8 & \\
	& & & & IV-I & 0.22 $\pm$ 0.15 & 1.5 & \\
	& & & & III-II & 0.19 $\pm$ 0.11 & 1.7 & \\ \cline{2-8} \vspace*{-3mm}\\
	& \multirow{8}{*}{red wing} & Q 0142-100 & A & II-I & -0.20 $\pm$ 0.08 & 2.5 & Intrinsic Variability? \\  \cline{3-8} \vspace*{-3mm}\\
	& & HE 0435-1223 & C & III-I & -0.49 $\pm$ 0.34 & 1.5 & \\  \cline{3-8} \vspace*{-3mm}\\
	& & SDSS 1004+4112 & A & III-I & 0.26 $\pm$ 0.08 & 3.1 & Microlensing Variability \\  \cline{3-8} \vspace*{-3mm}\\
	& & \multirow{5}{*}{HE 1104-1805} & \multirow{2}{*}{A} & II-I & 0.14 $\pm$ 0.06 & 2.1 & Intrinsic Variability \\
	& & & & IV-II & 0.11 $\pm$ 0.04 & 2.7 & Intrinsic Variability? \\ \cline{4-8} \vspace*{-3mm}\\
	& & & \multirow{3}{*}{B} & IV-I & 0.43 $\pm$ 0.20 & 2.2 & Intrinsic Variability \\
	& & & & IV-II & 0.35 $\pm$ 0.19 & 1.8 & \\
	& & & & IV-III & 0.26 $\pm$ 0.16 & 1.7 & \\
	\hline \vspace*{-3mm}\\
	C III] & red wing & HE 0047-1756 & A & III-II & -0.30 $\pm$ 0.20 & 1.5 & \\
	\hline \vspace*{-3mm}\\
	\multirow{6}{*}{Mg II} & \multirow{5}{*}{blue wing} & \multirow{3}{*}{HE 0435-1223} & A & IV-III & -0.12 $\pm$ 0.08 & 1.5 & \\ \cline{4-8} \vspace*{-3mm}\\
	& & & \multirow{2}{*}{D} & III-II & -0.21 $\pm$ 0.11 & 1.9 & \\
	& & & & IV-III & -0.20 $\pm$ 0.12 & 1.6 & \\  \cline{3-8} \vspace*{-3mm}\\
	& & FBQ 0951+2635 & A & II-I & -0.11 $\pm$ 0.07 & 1.5 & \\  \cline{3-8} \vspace*{-3mm}\\
	& & WFI 2033-4723 & C & III-II & 0.12 $\pm$ 0.07 & 1.6 & \\ \cline{2-8} \vspace*{-3mm}\\
	& red wing & HE 0435-1223 & D & IV-III & -0.37 $\pm$ 0.23 & 1.6 & \\
\enddata
\tablenotetext{*}{\small magnitude difference}
\tablenotetext{**}{\small standard deviation of magnitude difference}
\end{deluxetable*}
\label{1.5em_epochs}

\begin{deluxetable*}{lcccccc}
\tablewidth{\textwidth}
\renewcommand{\arraystretch}{0.75}
\tablecaption{Differences between Images - other Emission Line Features}
\tablehead{\colhead{Emission Line} & \colhead{Object} & \colhead{Image Pair} & \colhead{Epoch} & \colhead{d$^{*}$ $\pm$ $\sigma^{**}$} &  \colhead{d/$\sigma$} & \colhead{Classification}\vspace*{-1mm}} 
\startdata
	\multirow{10}{*}{red shelf $\lambda\lambda1580-1620$} & \multirow{3}{*}{HE 0435-1223} & B-A & IV & 0.25 $\pm$ 0.08 & 3.3 & Microlensing \\ \cline{3-7} \vspace*{-3mm}\\
	& & C-A & IV & 0.14 $\pm$ 0.08 & 1.7 & \\ \cline{3-7} \vspace*{-3mm}\\
	& & D-B & IV & -0.17 $\pm$ 0.06 & 2.9 & Microlensing \\ \cline{2-7} \vspace*{-3mm}\\
	& QSO 0957+561 & B-A & I & -0.37 $\pm$ 0.24 & 1.6 & \\ \cline{2-7} \vspace*{-3mm}\\
	& \multirow{3}{*}{SDSS J1004+4112} & C-A & I & 0.68 $\pm$ 0.34 & 2.0 & Microlensing \\ \cline{3-7} \vspace*{-3mm}\\
	& & D-A & I & 0.58 $\pm$ 0.23 & 2.5 & Microlensing \\ \cline{3-7} \vspace*{-3mm}\\
	& & C-B & I & 0.51 $\pm$ 0.11 & 4.6 & Microlensing \\ \cline{2-7} \vspace*{-3mm}\\
	& HE 1104-1805 & B-A & I & 0.15 $\pm$ 0.10 & 1.5 & \\ \cline{2-7} \vspace*{-3mm}\\
	& \multirow{3}{*}{SDSS 1339+1310} & \multirow{3}{*}{B-A} & I & -0.89 $\pm$ 0.48 & 1.9 & \\
	& &  & II & 1.53 $\pm$ 0.29 & 5.2 & Microlensing \\
	& &  & III & -0.77 $\pm$ 0.09 & 8.8 & Microlensing \\
	\hline \vspace*{-3mm}\\
	\multirow{4}{*}{Fe III $\lambda\lambda1978-2018$} & FBQS J0951+2635 & B-A & I & -1.11 $\pm$ 0.43 & 2.6 & Microlensing \\ \cline{2-7} \vspace*{-3mm}\\
	& QSO 0957+561 & B-A & II & -0.43 $\pm$ 0.11 & 3.9 & Microlensing  or Intrinsic Variability + $\Delta \tau $\\ \cline{2-7} \vspace*{-3mm}\\
	& \multirow{2}{*}{SDSS 1339+1310} &  \multirow{2}{*}{B-A} & I & -1.05 $\pm$ 0.71 & 1.5 & \\
	& & & II & -0.97 $\pm$ 0.56 & 1.7 & \\ 
	\hline \vspace*{-3mm}\\
	\multirow{4}{*}{Fe III $\lambda\lambda2039-2113$} & QSO 0957+561 & B-A & II & 0.40 $\pm$ 0.10 & 3.2 & Microlensing or Intrinsic Variability + $\Delta \tau $ \\ \cline{2-7} \vspace*{-3mm}\\
	& SDSS J1004+4112 & C-A & I & 1.29 $\pm$ 0.85 & 1.5 & \\ \cline{2-7} \vspace*{-3mm}\\
	& \multirow{2}{*}{SDSS 1339+1310} &  \multirow{2}{*}{B-A} & I & -0.96 $\pm$ 0.29 & 3.3 & Microlensing\\
	& & & II & -0.74 $\pm$ 0.37 & 2.0 & Microlensing \\ 
	\hline \vspace*{-3mm}\\
	\multirow{11}{*}{Fe III $\lambda\lambda2386-2449$} & HE 0047-1756 & B-A & II & -0.11 $\pm$ 0.07 & 1.6 & \\ \cline{2-7} \vspace*{-3mm}\\
	& \multirow{6}{*}{HE 0435-1223} & B-A & IV & -0.27 $\pm$ 0.13 & 2.1 & Microlensing \\ \cline{3-7} \vspace*{-3mm}\\
	& & C-A & III & -0.25 $\pm$ 0.16 & 1.6 & \\ \cline{3-7} \vspace*{-3mm}\\
	& & D-A & III & 0.19 $\pm$ 0.12 & 1.5 & \\ \cline{3-7} \vspace*{-3mm}\\
	& & C-B & IV & 0.29 $\pm$ 0.16 & 1.8 & \\ \cline{3-7} \vspace*{-3mm}\\
	& & D-B & IV & 0.28 $\pm$ 0.14 & 1.9 & \\ \cline{3-7} \vspace*{-3mm}\\
	& & D-C & III & 0.43 $\pm$ 0.23 & 1.9 & \\ \cline{2-7} \vspace*{-3mm}\\
	& SDSS J0806+2006 & B-A & I & -0.68 $\pm$ 0.24 & 2.8 & Microlensing \\ \cline{2-7} \vspace*{-3mm}\\
	& FBQS J0951+2635 & B-A & II & -0.19 $\pm$ 0.12 & 1.6 & \\ \cline{2-7} \vspace*{-3mm}\\
	& \multirow{2}{*}{SDSS 1339+1310} &  \multirow{2}{*}{B-A} & I & -0.42 $\pm$ 0.24 & 1.7 & \\
	& & & II & -0.30 $\pm$ 0.12 & 2.5 & Microlensing \\ 
	\hline \vspace*{-3mm}\\
	\multirow{4}{*}{Fe II $\lambda\lambda2158-2197$} & FBQS J0951+2635 & B-A & II & -0.70 $\pm$  0.27 & 2.6 & Microlensing \\ \cline{2-7} \vspace*{-3mm}\\
	& QSO 0957+561 & B-A & II & 0.37 $\pm$ 0.18 & 2.0 & Microlensing or Intrinsic Variability + $\Delta \tau $ \\ \cline{2-7} \vspace*{-3mm}\\
	& \multirow{2}{*}{SDSS 1339+1310} &  \multirow{2}{*}{B-A} & I & -1.24 $\pm$ 0.27 & 4.6 & Microlensing \\
	& & & II & -1.13 $\pm$ 0.72 & 1.6 & \\ 
	\hline \vspace*{-3mm}\\
	\multirow{1}{*}{Fe II $\lambda\lambda2209-2239$} & SDSS 1339+1310 & B-A & I & -1.62 $\pm$ 0.81 & 2.0 & Microlensing \\
	\hline \vspace*{-3mm}\\
	\multirow{4}{*}{Fe II $\lambda\lambda2261-2364$} & HE 0435-1223 & D-C & III & -0.23 $\pm$ 0.15 & 1.5 &  \\ \cline{2-7} \vspace*{-3mm}\\
	& FBQS J0951+2635 & B-A & I & -0.66 $\pm$ 0.37 & 1.8 & \\ \cline{2-7} \vspace*{-3mm}\\
	& \multirow{2}{*}{SDSS 1339+1310} &  \multirow{2}{*}{B-A} & I & -0.86 $\pm$ 0.30 & 2.9 & Microlensing \\
	& & & II & -0.46 $\pm$ 0.20 & 2.3 & Microlensing \\ 
	\hline \vspace*{-3mm}\\
	\multirow{9}{*}{Fe II $\lambda\lambda2460-2564$} & HE 0047-1756 & B-A & I & -0.12 $\pm$ 0.08 & 1.6 & \\ \cline{2-7} \vspace*{-3mm}\\
	& \multirow{5}{*}{HE 0435-1223} & B-A & IV & -0.33 $\pm$ 0.29 & 1.6 & \\ \cline{3-7} \vspace*{-3mm}\\
	& & C-A & III & -0.29 $\pm$ 0.12 & 2.5 & Microlensing \\ \cline{3-7} \vspace*{-3mm}\\
	& & D-A & III & 0.17 $\pm$  0.11 & 1.5 & \\ \cline{3-7} \vspace*{-3mm}\\
	& & C-B & IV & 0.33 $\pm$ 0.22 & 1.5 & \\ \cline{3-7} \vspace*{-3mm}\\
	& & D-C & III & 0.47 $\pm$ 0.13 & 3.6 & Microlensing \\ \cline{2-7} \vspace*{-3mm}\\
	& SDSS J0806+2006 & B-A & I & -0.72 $\pm$ 0.21 & 3.4 & Microlensing \\ \cline{2-7} \vspace*{-3mm}\\
	& FBQS J0951+2635 & B-A & II & -0.20 $\pm$ 0.09 & 2.2 & Microlensing \\ \cline{2-7} \vspace*{-3mm}\\
	& SDSS 1339+1310 & B-A & I & -0.16 $\pm$ 0.10 & 1.6 & \\ 
	\hline \vspace*{-3mm}\\
	\multirow{4}{*}{Fe II $\lambda\lambda2596-2645$} & \multirow{2}{*}{HE 0435-1223} & B-A & I & 1.28 $\pm$ 0.80 & 1.6 &\\ \cline{3-7} \vspace*{-3mm}\\
	& & C-B & I & -1.04 $\pm$ 0.68 & 1.5 & \\ \cline{2-7} \vspace*{-3mm}\\
	& FBQS J0951+2635 & B-A & II & -0.27 $\pm$ 0.17 & 1.6 & \\ \cline{2-7} \vspace*{-3mm}\\
	& SDSS J1004+4112 & B-A & I & -0.70 $\pm$ 0.38 & 1.9 & \\
\enddata
\tablenotetext{*}{magnitude difference}
\tablenotetext{**}{standard deviation of magnitude difference}
\end{deluxetable*}
\label{1.5o_images}

\clearpage
\startlongtable
\begin{deluxetable*}{lcccccc}
\tablewidth{\textwidth}
\setlength{\tabcolsep}{10pt}
\renewcommand{\arraystretch}{0.78}
\tablecaption{Differences between Epochs - other Emission Line Features}
\tablehead{\colhead{Emission Line} & \colhead{Object} & \colhead{Image} & \colhead{Epoch} & \colhead{d$^{*}$ $\pm$ $\sigma^{**}$} &  \colhead{d/$\sigma$} & \colhead{Classification}} 
\startdata
 	\multirow{18}{*}{red shelf $\lambda\lambda1580-1620$} & \multirow{1}{*}{Q 0142-100} & A & II-I & -0.36 $\pm$ 0.08 & 4.5 & Intrinsic Variability? \\ \cline{2-7} \vspace*{-3mm}\\
	& \multirow{5}{*}{HE 0435-1223} & \multirow{2}{*}{A} & III-I & -0.99 $\pm$ 0.52 & 1.9 & \\ 
	& & & IV-III & -0.32 $\pm$ 0.12 & 2.6 & Microlensing Variability \\ \cline{3-7} \vspace*{-3mm}\\
	& & B & IV-I & -0.48 $\pm$ 0.32 & 1.5 &  \\ \cline{3-7} \vspace*{-3mm}\\
	& & C & IV-I & -0.90 $\pm$ 0.60 & 1.5 & \\ \cline{3-7} \vspace*{-3mm}\\
	& & D & IV-I & -0.50 $\pm$ 0.30 & 1.6 & \\ \cline{2-7} \vspace*{-3mm}\\ 
	& \multirow{2}{*}{QSO 0957+561} & A & II-I & 0.50 $\pm$ 0.21 & 2.4& Intrinsic Variability? \\ \cline{3-7} \vspace*{-3mm}\\
	& & B & II-I & 0.93 $\pm$ 0.55 & 1.7 & \\ \cline{2-7} \vspace*{-3mm}\\
	& \multirow{2}{*}{SDSS J1004+4112} & \multirow{2}{*}{A} & II-I & 0.33 $\pm$ 0.18 & 1.9 & \\
	& & & III-I & 0.44 $\pm$ 0.08 & 5.4 & Microlensing Variability\\ \cline{2-7} \vspace*{-3mm}\\
	& \multirow{4}{*}{HE 1104-1805} &  \multirow{4}{*}{A} & III-I & 1.19 $\pm$ 0.67 & 1.8 & \\
	& & & IV-I & 0.33 $\pm$ 0.08 & 4.3 & Intrinsic Variability? \\ 
	& & & III-II & 1.24 $\pm$ 0.81 & 1.5 & \\
	& & & IV-II & 0.24 $\pm$ 0.08 & 2.9 & Intrinsic Variability? \\ \cline{2-7} \vspace*{-3mm}\\
	& SDSS 1339+1310 & A & III-II & 0.68 $\pm$ 0.24 & 2.8 & Microlensing Variability \\ \cline{2-7} \vspace*{-3mm}\\
	& \multirow{3}{*}{HE 2149-2745} & \multirow{2}{*}{A} & II-I & 0.52 $\pm$ 0.18 & 2.9 & Intrinsic Variability? \\
	& & & III-II & -0.51 $\pm$ 0.16 & 3.2 & Intrinsic Variability? \\ \cline{3-7} \vspace*{-3mm}\\
	& & B & II-I & 0.28 $\pm$ 0.19 & 1.5 & \\ 
	\hline \vspace*{-3mm}\\
	\multirow{3}{*}{Fe II $\lambda\lambda1705-1730$} & HE 1104-1805 & B & IV-III & -0.70 $\pm$ 0.53 & 1.5 & \\ \cline{2-7} \vspace*{-3mm}\\
	& \multirow{2}{*}{SDSS 1339+1310} & A & III-I & 1.24 $\pm$ 0.85 & 1.5 & \\ \cline{3-7} \vspace*{-3mm}\\
	& & B & III-I & 1.28 $\pm$ 0.74 & 1.7 & \\
	\hline \vspace*{-3mm}\\
	Fe II $\lambda\lambda1760-1800$ & Q 0142-100 & A & II-I & -1.17 $\pm$ 0.62 & 1.9 & \\
	\hline \vspace*{-3mm}\\
	\multirow{3}{*}{Fe III $\lambda\lambda1978-2018$} & QSO 0957+561 & A & II-I & 0.97 $\pm$ 0.55 & 1.8 & \\ \cline{2-7} \vspace*{-3mm}\\
	& HE 1104-1805 & A & III-II & -0.73 $\pm$ 0.40 & 1.8 & \\ \cline{2-7} \vspace*{-3mm}\\
	& \multirow{1}{*}{SDSS 1339+1310} & B & II-I & 1.02 $\pm$ 0.67 & 1.5 & \\ 
	\hline \vspace*{-3mm}\\
	\multirow{2}{*}{Fe III $\lambda\lambda2039-2113$} & \multirow{2}{*}{QSO 0957+561} & A & II-I & 1.30 $\pm$ 0.67 & 1.9 & \\ \cline{3-7} \vspace*{-3mm}\\
	& & B & II-I & 1.70 $\pm$ 0.89 & 1.9 & \\ 
	\hline \vspace*{-3mm}\\
	\multirow{12}{*}{Fe III $\lambda\lambda2386-2449$} & \multirow{2}{*}{HE 0047-1756} & A & II-I & -0.82 $\pm$ 0.08 & 10.4 & Intrinsic Variability  \\ \cline{3-7} \vspace*{-3mm}\\
	& & B & II-I & -0.97 $\pm$ 0.12 & 8.4 & Intrinsic Variability \\ \cline{2-7} \vspace*{-3mm}\\
	& \multirow{4}{*}{HE 0435-1223} & A & IV-III & 0.43 $\pm$ 0.17 & 2.6 & Intrinsic Variability\\ \cline{3-7} \vspace*{-3mm}\\
	& & B & IV-I & -0.39 $\pm$ 0.17 & 2.4 & Microlensing Variability \\ \cline{3-7} \vspace*{-3mm}\\
	& & \multirow{2}{*}{C} & III-I & -0.58 $\pm$ 0.28 & 2.1 & Intrinsic Variability? \\
	& & & IV-III & 0.70 $\pm$ 0.16 & 4.4 & Intrinsic Variability \\ \cline{2-7} \vspace*{-3mm}\\
	& SDSS J0806+2006 & B & II-I & 0.78 $\pm$ 0.40 & 2.0 & Microlensing Variability \\ \cline{2-7} \vspace*{-3mm}\\
	& SDSS J1004+4112 & A & III-I & -0.43 $\pm$ 0.20 & 2.2 & Microlensing Variability? \\ \cline{2-7} \vspace*{-3mm}\\
	&  \multirow{2}{*}{WFI J2033-4723} & B & II-I & -0.25 $\pm$ 0.13 & 2.0 & Intrinsic Variability \\ \cline{3-7} \vspace*{-3mm}\\
	& & C & II-I & -0.31 $\pm$ 0.11 & 2.7 & Intrinsic Variability \\ \cline{2-7} \vspace*{-3mm}\\
	& \multirow{2}{*}{HE 2149-2745} & \multirow{2}{*}{B} & II-I & 0.26 $\pm$ 0.13 & 2.0 & Intrinsic Variability? \\
	& & & III-II & -0.28 $\pm$ 0.12 & 2.4 & Intrinsic Variability? \\
	\hline \vspace*{-3mm}\\
	\multirow{1}{*}{Fe II $\lambda\lambda2158-2197$} & QSO 0957+561 & B & II-I & 1.13 $\pm$ 0.73 & 1.6 & \\ 
	\hline \vspace*{-3mm}\\
	\multirow{3}{*}{Fe II $\lambda\lambda2209-2239$} & \multirow{3}{*}{HE 0435-1223} & \multirow{2}{*}{C} & IV-I & 0.69 $\pm$ 0.42 & 1.7 & \\
	& & & IV-III & 1.34 $\pm$ 0.86 & 1.6 & \\ \cline{3-7} \vspace*{-3mm}\\
	& & D & III-II & -1.14 $\pm$ 0.53 & 2.1 & Intrinsic Variability?\\ 
	\hline \vspace*{-3mm}\\
	\multirow{8}{*}{Fe II $\lambda\lambda2261-2364$} & \multirow{2}{*}{HE 0047-1756} & A & II-I & -0.46 $\pm$ 0.13 & 3.5 & Intrinsic Variability? \\ \cline{3-7} \vspace*{-3mm}\\
	& & B & II-I & -0.71 $\pm$ 0.46 & 1.5 & \\ \cline{2-7} \vspace*{-3mm}\\
	& \multirow{5}{*}{HE 0435-1223} & \multirow{2}{*}{A} & IV-I & 0.75 $\pm$ 0.46 & 1.6 & \\ 
	& & & IV-III & 0.88 $\pm$ 0.41 & 2.1 & Intrinsic Variability \\  \cline{3-7} \vspace*{-3mm}\\
	& & C & IV-III & 0.84 $\pm$ 0.34 & 2.5 & Intrinsic Variability  \\ \cline{3-7} \vspace*{-3mm}\\
	& & \multirow{2}{*}{D} & III-II & -0.60 $\pm$ 0.22 & 2.8 & Intrinsic Variability? \\ 
	& & & IV-III & 1.09 $\pm$ 0.54 & 2.0 & Intrinsic Variability \\ \cline{2-7} \vspace*{-3mm}\\
	& FBQS J0951+2635 & B & II-I & 0.52 $\pm$ 0.24 & 2.2 & Microlensing Variability \\
	\hline \vspace*{-3mm}\\
	\multirow{16}{*}{Fe II $\lambda\lambda2460-2564$} & \multirow{2}{*}{HE 0047-1756} & A & II-I & -0.74 $\pm$ 0.07 & 11.3 & Intrinsic Variability \\ \cline{3-7} \vspace*{-3mm}\\
	& & B & II-I & -0.87 $\pm$ 0.08 & 10.5 & Intrinsic Variability \\ \cline{2-7} \vspace*{-3mm}\\
	& \multirow{5}{*}{HE 0435-1223} & \multirow{2}{*}{A} & IV-I & 0.49 $\pm$ 0.32 & 1.5 & \\
	& &  & IV-III & 0.65 $\pm$ 0.21 & 3.1 & Intrinsic Variability \\ \cline{3-7} \vspace*{-3mm}\\
	& & \multirow{2}{*}{C} & III-I & -0.57 $\pm$  0.24 & 2.4 & Microlensing Variability\\
	& &  & IV-III & 0.95 $\pm$ 0.20 & 4.7 & Intrinsic Variability \\ \cline{3-7} \vspace*{-3mm}\\
	& & D & IV-II & 0.44 $\pm$ 0.28 & 1.6 & \\ \cline{2-7} \vspace*{-3mm}\\
	& SDSS J0806+2006 & A & II-I & 0.61 $\pm$ 0.31 & 1.9 & \\ \cline{2-7} \vspace*{-3mm}\\
	& \multirow{2}{*}{FBQS J0951+2635} & A & II-I & -0.19 $\pm$ 0.12 & 1.5 & \\ \cline{3-7} \vspace*{-3mm}\\
	& & B & II-I & -0.27 $\pm$ 0.18 & 1.5 & \\ \cline{2-7} \vspace*{-3mm}\\
	& SDSS 1339+1310 & A & II-I & -0.30 $\pm$ 0.19 & 1.6 & \\ \cline{2-7} \vspace*{-3mm}\\
	& WFI J2033-4723 & C & II-I & -0.24 $\pm$ 0.10 & 2.5 & Intrinsic Variability? \\ \cline{2-7} \vspace*{-3mm}\\
	& \multirow{4}{*}{HE 2149-2745} & \multirow{2}{*}{A} & II-I & 0.28 $\pm$ 0.09 & 3.0 & Intrinsic Variability \\
	& & & III-II & -0.29 $\pm$ 0.09 & 3.3 & Intrinsic Variability  \\ \cline{3-7} \vspace*{-3mm}\\
	& & \multirow{2}{*}{B} & II-I & 0.37 $\pm$ 0.15 & 2.5 & Intrinsic Variability \\
	& & & III-II & -0.38 $\pm$ 0.16 & 2.4 & Intrinsic Variability \\
	\hline \vspace*{-3mm}\\
	\multirow{8}{*}{Fe II $\lambda\lambda2596-2645$} & \multirow{2}{*}{HE 0047-1756} & A & II-I & -0.68 $\pm$ 0.17 & 4.0 & Intrinsic Variability \\ \cline{3-7} \vspace*{-3mm}\\
	& & B & II-I & -0.82 $\pm$ 0.22 & 3.7 & Intrinsic Variability \\ \cline{2-7} \vspace*{-3mm}\\
	& \multirow{6}{*}{HE 0435-1223} & \multirow{2}{*}{A} & IV-I & 1.40 $\pm$ 0.77 & 1.8 & \\
	& & & IV-III & 1.34 $\pm$ 0.78 & 1.7 & \\ \cline{3-7} \vspace*{-3mm}\\
	& & B & IV-II & 0.98 $\pm$ 0.47 & 2.1 & Intrinsic Variability \\ \cline{3-7} \vspace*{-3mm}\\
	& & C & IV-III & 1.28 $\pm$ 0.71 & 1.8 & \\ \cline{3-7} \vspace*{-3mm}\\
	& & \multirow{2}{*}{D} & IV-II & 1.16 $\pm$ 0.43 & 2.7 & Intrinsic Variability \\
	& & & IV-III & 0.97 $\pm$ 0.57 & 1.7 & \\
\enddata
\tablenotetext{*}{magnitude difference}
\tablenotetext{**}{standard deviation of magnitude difference}
\end{deluxetable*}
\label{1.5o_epochs}

\begin{table*}
\centering
\caption{Core differences between epochs}
\tabcolsep=0.3cm
\renewcommand{\arraystretch}{0.8}
\begin{tabular}{lcccc|cc}
	\hline \hline \vspace*{-3mm}\\
	\multirow{2}{*}{Object} & \multirow{2}{*}{Image Pair} & \multirow{2}{*}{Epoch} & \multicolumn{2}{c|}{C IV} & \multicolumn{2}{c}{C III} \\
	\cline{4-7} \vspace*{-3mm}\\
	& & & d$^{*}$ $\pm$ $\sigma^{**}$ & d/$\sigma$ & d$^{*}$ $\pm$ $\sigma^{**}$ &  d/$\sigma$  \\
	\hline \vspace*{-3mm}\\
	\multirow{2}{*}{HE 0047-1756}  & A & III-II & \dots & \dots & -0.01 $\pm$ 0.05 & 0.2 \\ \cline{2-7} \vspace*{-3mm}\\
	& B & III-II & \dots & \dots & 0.02 $\pm$ 0.04 & 0.5 \\
	\hline \vspace*{-3mm}\\
	\multirow{2}{*}{Q 0142-100} & A & II-I & -0.55 $\pm$ 0.03 & 18.1 & \dots & \dots \\ \cline{2-7} \vspace*{-3mm}\\
	& B & II-I & -0.53 $\pm$ 0.03 & 16.2 & \dots & \dots \\
	\hline \vspace*{-3mm}\\
	\multirow{13}{*}{HE 0435-1223} & \multirow{3}{*}{A} & III-I & 0.85 $\pm$ 0.05 & 16.0 & \dots & \dots \\
	&  & IV-I & 0.64 $\pm$ 0.03 & 21.4 & \dots & \dots \\
	&  & IV-III & -0.21 $\pm$ 0.06 & 3.4 & -0.73 $\pm$ 0.06 & 11.6 \\ \cline{2-7} \vspace*{-3mm}\\
	& \multirow{2}{*}{B} & IV-I & 0.35 $\pm$ 0.05 & 7.7 & \dots & \dots \\
	&  & IV-II & \dots & \dots & -0.39 $\pm$ 0.02 & 24.6 \\ \cline{2-7} \vspace*{-3mm}\\
	& \multirow{3}{*}{C} & III-I & 0.74 $\pm$ 0.04 & 20.9 & \dots & \dots \\
	&  & IV-I & 0.67 $\pm$ 0.08 & 8.3 & \dots & \dots \\
	&  & IV-III & -0.09 $\pm$ 0.08 & 1.1 & -0.67 $\pm$ 0.05 & 13.7 \\ \cline{2-7} \vspace*{-3mm}\\
	& \multirow{5}{*}{D} & III-I & 1.02 $\pm$ 0.08 & 13.5 & \dots & \dots \\
	& & IV-I & 0.33 $\pm$ 0.08 & 4.1 & \dots & \dots \\
	& & III-II & \dots & \dots & 0.03 $\pm$ 0.06 & 0.5 \\
	& & IV-II & \dots & \dots & -0.44 $\pm$ 0.06 & 7.1 \\ 
	& & IV-III & -0.69 $\pm$ 0.12 & 5.9 & -0.47 $\pm$ 0.08 & 6.1 \\ 
	\hline \vspace*{-3mm}\\
	\multirow{2}{*}{SDSS J0806+2006} & A & II-I & \dots & \dots & 0.24 $\pm$ 0.07 & 3.5 \\ \cline{2-7} \vspace*{-3mm}\\
	& B & II-I & \dots & \dots & 0.46 $\pm$  0.09 & 4.9 \\
    \hline \vspace*{-3mm}\\
    \multirow{2}{*}{QSO 0957+561} & A & II-I & -0.49 $\pm$ 0.09 & 5.2 & 0.12 $\pm$ 0.03 & 3.8 \\ \cline{2-7} \vspace*{-3mm}\\
    & B & II-I & -0.33 $\pm$ 0.04 & 8.8 & 0.17 $\pm$ 0.07 & 2.4 \\
    \hline \vspace*{-3mm}\\
    \multirow{6}{*}{SDSS J1004+4112} & \multirow{3}{*}{A} & II-I & -0.08 $\pm$ 0.02 & 3.5 & \dots & \dots \\
    &  & III-I & 0.14 $\pm$ 0.05 & 3.1 & -0.04 $\pm$ 0.04 & 1.2 \\ 
    &  & III-II & 0.24 $\pm$ 0.07 & 3.3 & \dots & \dots  \\ \cline{2-7} \vspace*{-3mm}\\
    & \multirow{3}{*}{B} & II-I & 0.02 $\pm$ 0.05 & 0.3 & \dots & \dots \\
    &  & III-I & 0.20 $\pm$ 0.04 & 4.6 & -0.15 $\pm$ 0.02 & 7.8 \\ 
    &  & III-II & 0.18 $\pm$ 0.07 & 2.4 & \dots & \dots \\
    \hline \vspace*{-3mm}\\
    \multirow{12}{*}{HE 1104-1805} & \multirow{6}{*}{A} & II-I & 0.37 $\pm$ 0.01 & 27.2 & \dots & \dots \\
     &  & III-I & 0.44 $\pm$ 0.05 & 9.2 & \dots & \dots \\
     &  & IV-I & 0.51 $\pm$ 0.03 & 15.5 & \dots & \dots \\
     &  & III-II & 0.06 $\pm$ 0.03 & 1.9 & \dots & \dots \\
     &  & IV-II & 0.27 $\pm$ 0.01 & 21.4 & \dots & \dots \\
     &  & IV-III & 0.21 $\pm$ 0.04 & 5.2 & \dots & \dots \\ \cline{2-7} \vspace*{-3mm}\\
     & \multirow{6}{*}{B} & II-I & 0.33 $\pm$ 0.02 & 18.0 & \dots & \dots \\
     &  & III-I & -0.04 $\pm$ 0.03 & 1.3 & \dots & \dots \\
     &  & IV-I & 0.45 $\pm$ 0.04 & 10.6 & \dots & \dots \\
     &  & III-II & -0.04 $\pm$ 0.03 & 1.3 & \dots & \dots \\
     &  & IV-II & 0.12 $\pm$ 0.05 & 2.7 & \dots & \dots \\
     &  & IV-III & 0.17 $\pm$ 0.04 & 4.7 & \dots & \dots \\
     \hline \vspace*{-3mm}\\
     \multirow{6}{*}{SDSS 1339+1310} & \multirow{3}{*}{A} & II-I & 0.11 $\pm$ 0.04 & 2.9  & -0.32 $\pm$ 0.07 & 22.5 \\ 
     &  & III-I & 0.03 $\pm$ 0.04 & 0.7 & \dots & \dots \\
     &  & III-II & -0.09 $\pm$ 0.02 & 4.0 & \dots & \dots \\ \cline{2-7} \vspace*{-3mm}\\
     & \multirow{3}{*}{B} & II-I & 0.13 $\pm$ 0.06 & 2.0 & -0.38 $\pm$  0.05 & 7.3 \\
     &  & III-I & 0.16 $\pm$ 0.03 & 5.2 & \dots & \dots \\
     &  & III-II & 0.03 $\pm$ 0.06 & 0.5 & \dots & \dots \\
     \hline \vspace*{-3mm}\\
     \multirow{2}{*}{WFI J2033-4723} & B & III-II & \dots & \dots & -0.03 $\pm$ 0.06 & 0.6 \\ \cline{2-7} \vspace*{-3mm}\\
     & C & III-II & \dots & \dots & -0.03 $\pm$ 0.04 & 0.9 \\
     \hline \vspace*{-3mm}\\
     \multirow{6}{*}{HE 2149-2745} & \multirow{3}{*}{A} & II-I & -0.17 $\pm$ 0.03 & 5.1 & -0.31 $\pm$ 0.03 & 10.2 \\
     & & III-I & -0.02 $\pm$ 0.01 & 1.4 & -0.04 $\pm$ 0.01 & 4.3 \\
     & & III-II & 0.13 $\pm$ 0.05 & 2.5 & 0.26 $\pm$ 0.03 & 9.6 \\ \cline{2-7} \vspace*{-3mm}\\
     & \multirow{3}{*}{B} & II-I & -0.23 $\pm$ 0.04 & 6.0 & -0.28 $\pm$ 0.05 & 5.3 \\
     & & III-I & -0.02 $\pm$ 0.02 & 1.2 & -0.04 $\pm$ 0.03 & 1.1 \\
     & & III-II & 0.21 $\pm$ 0.05 & 4.5 & 0.24 $\pm$ 0.06 & 4.2 \\
     \hline \vspace*{-3mm}\\
\end{tabular}
\begin{flushleft}
\tablenotetext{*}{magnitude difference}
\tablenotetext{**}{standard deviation of magnitude difference}
\end{flushleft}
\end{table*}
\label{corevar}

\begin{table*}
\centering
\tabcolsep=0.3cm
\renewcommand{\arraystretch}{0.8}
	\caption{Core differences between images}
	\begin{tabular}{lcccc|cc}
		\hline \hline \vspace*{-3mm}\\
		\multirow{2}{*}{Object} & \multirow{2}{*}{Image Pair} & \multirow{2}{*}{Epoch} & \multicolumn{2}{c|}{C IV} & \multicolumn{2}{c}{C III} \\
		\cline{4-7} \vspace*{-3mm}\\
		& & & d$^{*}$ $\pm$ $\sigma^{**}$ & d/$\sigma$ & d$^{*}$ $\pm$ $\sigma^{**}$ &  d/$\sigma$  \\
		\hline \vspace*{-3mm}\\
		\multirow{3}{*}{HE 0047-1756}  & \multirow{3}{*}{B-A} & I & -0.08 $\pm$  0.03 & 1.1 & \dots & \dots \\
		& & II & \dots & \dots & -0.01 $\pm$ 0.01 & 1.2 \\
		& & III & \dots & \dots & 0.01 $\pm$ 0.04 & 0.4\\
		\hline \vspace*{-3mm}\\
		\multirow{2}{*}{Q 0142-100} & \multirow{2}{*}{B-A} & I & 0.03 $\pm$ 0.03 & 1.1 & \dots & \dots \\
		& & II & 0.05 $\pm$ 0.02 & 2.2 & \dots & \dots \\
		\hline \vspace*{-3mm}\\
		\multirow{16}{*}{HE 0435-1223} & \multirow{2}{*}{B-A} & I & -0.07 $\pm$ 0.04 & 1.7 & \dots & \dots \\ 		
		& & IV & -0.34 $\pm$ 0.03 & 12.3 & 0.09 $\pm$ 0.02 & 4.1 \\ \cline{2-7} \vspace*{-3mm}\\
		& \multirow{3}{*}{C-A} & I & 0.02 $\pm$ 0.07 & 0.3 & \dots & \dots \\
		& & III & -0.08 $\pm$ 0.04 & 2.0 & 0.08 $\pm$ 0.04 & 2.1 \\
		& & IV & 0.05 $\pm$ 0.04 & 1.2 & 0.14 $\pm$ 0.02 & 7.2 \\ \cline{2-7} \vspace*{-3mm}\\
		& \multirow{3}{*}{D-A} & I & -0.09 $\pm$ 0.03 & 2.7 & \dots & \dots \\
		& & III & 0.08 $\pm$ 0.07 & 1.2 & -0.19 $\pm$ 0.07 & 2.9 \\
		& & IV & -0.39 $\pm$ 0.03 & 13.2 & 0.07 $\pm$ 0.05 & 1.4 \\ \cline{2-7} \vspace*{-3mm}\\
		& \multirow{2}{*}{C-B} & I & 0.08 $\pm$ 0.08 & 1.0 & \dots & \dots \\
		& & IV & 0.38 $\pm$ 0.04 & 10.8 & 0.06 $\pm$ 0.02 & 3.3 \\ \cline{2-7} \vspace*{-3mm}\\
		& \multirow{3}{*}{D-B} & I & -0.03 $\pm$ 0.05 & 0.6 & \dots & \dots \\
		& & II & \dots & \dots & 0.03 $\pm$ 0.03 & 1.2 \\
		& & IV & -0.05 $\pm$ 0.03 & 1.5 & -0.01 $\pm$ 0.03 & 0.3 \\ \cline{2-7} \vspace*{-3mm}\\
		& \multirow{3}{*}{D-C} & I & -0.11 $\pm$ 0.03 & 3.2 & \dots & \dots \\
		& & III & 0.16 $\pm$ 0.09 & 1.8 & -0.26 $\pm$ 0.08 & 3.5 \\
		& & IV & -0.44 $\pm$ 0.02 &19.6 & -0.07 $\pm$ 0.04 & 1.8 \\
		\hline \vspace*{-3mm}\\
		\multirow{2}{*}{SDSS J0806+2006} & \multirow{2}{*}{B-A} & I & \dots & \dots & -0.23 $\pm$ 0.03 & 7.5 \\
		& & II & \dots & \dots & -0.02 $\pm$  0.05 & 0.3 \\
		\hline \vspace*{-3mm}\\
		FBQS J0951+2635 & B-A & I & \dots & \dots & 0.25 $\pm$ 0.08 & 3.3 \\
		\hline \vspace*{-3mm}\\
		\multirow{2}{*}{QSO 0957+561} & \multirow{2}{*}{B-A} & I & -0.09 $\pm$ 0.05 & 2.0 & 0.10 $\pm$ 0.02 & 5.3 \\
		& & II & 0.08 $\pm$ 0.07 & 1.1 & 0.14 $\pm$ 0.04 & 3.3 \\
		\hline \vspace*{-3mm}\\
		\multirow{8}{*}{SDSS J1004+4112} & \multirow{3}{*}{B-A} & I & 0.06 $\pm$ 0.02 & 2.5 & 0.11 $\pm$ 0.03 & 3.4 \\
		& & II & 0.17 $\pm$ 0.06 & 2.6 & \dots & \dots \\
		& & III & 0.10 $\pm$ 0.04 & 2.3 & 0.00 $\pm$ 0.03 & 0.1 \\ \cline{2-7} \vspace*{-3mm}\\
		& C-A & I & 0.05 $\pm$ 0.04 & 1.1 & 0.18 $\pm$ 0.13 & 1.4 \\ \cline{2-7} \vspace*{-3mm}\\
		& D-A & I & 0.13 $\pm$ 0.03 & 4.8 & 0.19 $\pm$  0.04 & 4.6 \\ \cline{2-7} \vspace*{-3mm}\\
		& C-B & I & -0.01 $\pm$ 0.05 & 0.3 & 0.07 $\pm$ 0.13 & 0.5 \\ \cline{2-7} \vspace*{-3mm}\\
		& D-B & I & 0.08 $\pm$ 0.03 & 2.5 & 0.07 $\pm$  0.05 & 1.4 \\ \cline{2-7} \vspace*{-3mm}\\
		& D-C & I & 0.10 $\pm$ 0.03 & 3.5 & 0.00 $\pm$ 0.12 & 0.0 \\ 
		\hline \vspace*{-3mm}\\
		\multirow{4}{*}{HE 1104-1805} & \multirow{4}{*}{B-A} & I & 0.06 $\pm$ 0.02 & 3.0 & \dots & \dots \\    
		& & II & 0.01 $\pm$ 0.02 & 0.07 & -0.13 $\pm$ 0.14 & 1.0 \\
		& & III & -0.08 $\pm$ 0.03 & 2.6 & \dots & \dots \\
		& & IV & -0.14 $\pm$ 0.02 & 7.7 & \dots & \dots \\
		\hline \vspace*{-3mm}\\
		\multirow{3}{*}{SDSS 1339+1310} & \multirow{3}{*}{B-A} & I & -0.35 $\pm$ 0.04 & 9.9 & -0.06 $\pm$ 0.07 & 0.9 \\
		& & II & -0.33 $\pm$ 0.06 & 5.2 & -0.13 $\pm$  0.02 & 5.5 \\
		& & III & -0.20 $\pm$ 0.03 & 8.1 & \dots & \dots \\
		\hline \vspace*{-3mm}\\
		\multirow{8}{*}{WFI J2033-4723} & A2-A1 & I & -0.10 $\pm$ 0.16 & 0.6 & \dots & \dots \\ \cline{2-7} \vspace*{-3mm}\\
		& B-A1 & I & -0.26 $\pm$ 0.10 & 2.5 & \dots & \dots \\ \cline{2-7} \vspace*{-3mm}\\
		& B-A2 & I & -0.21 $\pm$ 0.11 & 1.9 & \dots & \dots \\  \cline{2-7} \vspace*{-3mm}\\
		& C-A1 & I & -0.16 $\pm$ 0.10 & 1.7 & \dots & \dots \\ \cline{2-7} \vspace*{-3mm}\\
		& C-A2 & I & -0.11 $\pm$ 0.12 & 1.0 & \dots & \dots \\ \cline{2-7} \vspace*{-3mm}\\
		& \multirow{3}{*}{C-B} & I & 0.08 $\pm$ 0.14 & 0.5 & \dots & \dots \\
		& & II & \dots & \dots & 0.03 $\pm$ 0.02 & 1.1 \\
		& & III &  \dots & \dots & 0.03 $\pm$ 0.02 & 1.4 \\
		\hline \vspace*{-3mm}\\
		\multirow{3}{*}{HE 2149-2745} & \multirow{3}{*}{B-A} & I & 0.04 $\pm$ 0.04 & 0.9 & -0.01 $\pm$ 0.03 & 0.3 \\
		& & II & -0.03 $\pm$ 0.04 & 0.8 & 0.03 $\pm$ 0.03 & 0.9 \\
		& & III & 0.03 $\pm$ 0.05 & 0.6 & 0.00 $\pm$ 0.04 & 0.0 \\
		\hline \vspace*{-3mm}\\
	\end{tabular}
\begin{flushleft}
\tablenotetext{*}{magnitude difference}
\tablenotetext{**}{standard deviation of magnitude difference}
\end{flushleft}    
\end{table*}
\label{extinction}

\end{document}